\documentclass[longauth]{aa}

\usepackage{amsmath}
\usepackage{graphicx}
\usepackage{graphics}
\usepackage{epsfig}
\usepackage{pdfpages} 
\usepackage{blindtext}
\usepackage{deluxetable}
\newcommand{\znii}{Zn\,{\sc ii}}
\newcommand{\feii}{Fe\,{\sc ii}}

\newcommand{\sii}{S\,{\sc ii}}
\newcommand{\hi}{H\,{\sc i}}
\newcommand{\ci}{C\,{\sc i}}

\usepackage[normalem]{ulem} 
\definecolor{darkgreen}{rgb}{0.0,0.75,0.0}

\usepackage{txfonts}
\usepackage[colorlinks=true,linkcolor=blue,citecolor=blue,urlcolor=blue]{hyperref}%
\begin{document}

\title{The cosmic build-up of dust and metals}
\subtitle{Accurate abundances from GRB-selected star-forming galaxies at $1.7 < z < 6.3$}
\titlerunning{Dust and metal abundances of GRB-selected star-forming galaxies at $1.7 < z < 6.3$}

\author{
K.~E.~Heintz,\inst{1,2}
A.~De~Cia,\inst{3,4}
C.~C.~Th\"one,\inst{5}
J.-K.~Krogager,\inst{6}
R.~M.~Yates,\inst{7}
S.~Vejlgaard,\inst{1,2}
C.~Konstantopoulou,\inst{4}
J.~P.~U.~Fynbo,\inst{1,2}
D.~Watson,\inst{1,2}
D.~Narayanan,\inst{8,1}
S.~N.~Wilson,\inst{1,2}
M.~Arabsalmani,\inst{9,10}
S.~Campana,\inst{11}
V.~D'Elia,\inst{12,13}
M.~De~Pasquale,\inst{14}
D.~H.~Hartmann,\inst{15}
L.~Izzo,\inst{16,17}
P.~Jakobsson,\inst{18}
C.~Kouveliotou,\inst{19}
A.~Levan,\inst{20}
Q.~Li,\inst{21}
D.~B.~Malesani,\inst{1,2,20}
A.~Melandri,\inst{11,13}
B.~Milvang-Jensen,\inst{1,2}
P.~M\o ller,\inst{3}
E.~Palazzi,\inst{22}
J.~Palmerio,\inst{23}
P.~Petitjean,\inst{24}
G.~Pugliese, \inst{25}
A.~Rossi,\inst{22}
A.~Saccardi,\inst{23}
R.~Salvaterra,\inst{26}
S.~Savaglio,\inst{27,22,28}
P.~Schady,\inst{29}
G.~Stratta,\inst{22,30,31,32}
N.~R.~Tanvir,\inst{33}
A.~de~Ugarte~Postigo,\inst{34}
S.~D.~Vergani,\inst{23, 24}
K.~Wiersema,\inst{35,7}
R.~A.~M.~J.~Wijers\inst{25}
T.~Zafar,\inst{36}
}
\institute{
Cosmic Dawn Center (DAWN), Denmark 
\and
Niels Bohr Institute, University of Copenhagen, Jagtvej 128, 2200 Copenhagen N, Denmark 
\and
European Southern Observatory, Karl-Schwarzschild Str. 2, 85748 Garching bei München, Germany 
\and 
Department of Astronomy, University of Geneva, Chemin Pegasi 51, 1290 Versoix, Switzerland 
\and
Astronomical Institute of the Czech Academy of Sciences (ASU-CAS), Fri\v cova 298, 251 65 Ond\v rejov, Czechia 
\and
Centre de Recherche Astrophysique de Lyon, CNRS, Univ. Claude Bernard Lyon 1, 9 Av. Charles André, 69230 Saint-Genis-Laval, France 
\and 
Centre for Astrophysics Research, University of Hertfordshire, Hatfield, AL10 9AB, UK 
\and
Department of Astronomy, University of Florida, Gainesville, FL 32611, USA  
\and
Excellence Cluster ORIGINS, Boltzmannstra{\ss}e 2, 85748 Garching, Germany 
\and
Ludwig-Maximilians-Universit\"at, Schellingstra{\ss}e 4, 80799 M\"unchen, Germany 
\and 
INAF - Osservatorio astronomico di Brera, Via Bianchi 46, Merate (LC), I-23807, Italy 
\and
Space Science Data Center (SSDC) - Agenzia Spaziale Italiana (ASI), 00133 Roma, Italy 
\and
INAF - Osservatorio Astronomico di Roma, Via Frascati 33, 00040 Monte Porzio Catone, Italy 
\and 
Mathematics, Informatics, Physics and Earth Science Department of Messina University, Papardo campus, Via F. S. D'Alcontres 31, 98166, Messina, Italy 
\and
Clemson University, Department of Physics and Astronomy, Clemson, SC 29634, USA 
\and
INAF - Osservatorio Astronomico di Capodimonte, Salita Moiariello 16, 80131, Napoli, Italy 
\and
DARK, Niels Bohr Institute, University of Copenhagen, Jagtvej 128, 2200 Copenhagen, Denmark 
\and
Centre for Astrophysics and Cosmology, Science Institute, University of Iceland, Dunhagi 5, 107 Reykjavik, Iceland 
\and
The George Washington University, Department of Physics, 725 21st street NW, Washington DC 20052, USA 
\and
Department of Astrophysics/IMAPP, Radboud University, 6525 AJ Nijmegen, The Netherlands 
\and
Max-Planck-Institut f{\"u}r Astrophysik, Karl-Schwarzschild-Str. 1, Garching b. M{\"u}nchen D-85741, Germany 
\and
INAF – Osservatorio di Astrofisica e Scienza dello Spazio, Via Piero Gobetti 93/3, 40129 Bologna, Italy 
\and
GEPI, Observatoire de Paris, Université PSL, CNRS, 5 Place Jules Janssen, 92190 Meudon, France 
\and
Institut d'Astrophysique de Paris and Sorbonne Universit\'e, 98bis Boulevard Arago, 75014, Paris 
\and
Anton Pannekoek Institute for Astronomy, University of Amsterdam, P.O. Box 94249, 1090GE Amsterdam, The Netherlands 
\and
Istituto Nazionale di Astrofisica (INAF) Istituto di Astrofisica Spaziale e Fisica Cosmica, Via Alfonso Corti 12, I-20133, Milano, Italy 
\and
Physics Department, University of Calabria, 87036 Arcavacata di Rende, CS, Italy 
\and
INFN – Laboratori Nazionali di Frascati, Frascati, Italy 
\and
Department of Physics, University of Bath, Claverton Down, Bath BA2 7AY, UK 
\and
Institut für Theoretische Physik, Johann Wolfgang Goethe-Universität, Max-von-Laue-Str. 1, 60438 Frankfurt am Main, Germany 
\and
Istituto di Astrofisica e Planetologia Spaziali di Roma, 00133 Roma, Italy 
\and
 INFN, Sezione di Roma, I-00185 Roma, Italy 
\and
School of Physics and Astronomy, University of Leicester, University Road, Leicester, LE1 7RH, UK 
\and 
Artemis, Observatoire de la Côte d’Azur, Université Côte d’Azur, CNRS, 06304 Nice, France 
\and
Physics Department, Lancaster University, Lancaster LA1 4YB, UK 
\and
School of Mathematical and Physical Sciences, Macquarie University, NSW 2109, Australia 
}
\authorrunning{Heintz et al.}

\date{Submitted ---, accepted ---, published ---}

\abstract{
The chemical enrichment of dust and metals in the interstellar medium (ISM) of galaxies throughout cosmic time is one of the key driving processes of galaxy evolution. Here we study the evolution of the gas-phase metallicities, dust-to-gas (DTG), and dust-to-metal (DTM) ratios of 36 star-forming galaxies at $1.7 < z < 6.3$ probed by gamma-ray bursts (GRBs). We compile all GRB-selected galaxies with intermediate ($\mathcal{R} = 7000$) to high ($\mathcal{R}>40,000$) resolution spectroscopic data, including three new sources, for which at least one refractory (e.g. Fe) and one volatile (e.g. S or Zn) element have been detected at S/N$>3$. This is to ensure that accurate abundances and dust depletion patterns can be obtained. We first derive the redshift evolution of the dust-corrected, absorption-line based gas-phase metallicity ${\rm [M/H]_{tot}}$ in these galaxies, for which we determine a linear relation with redshift ${\rm [M/H]_{tot}}(z) = (-0.21\pm 0.04)z -(0.47\pm 0.14)$. We then examine the DTG and DTM ratios as a function of redshift and through three orders of magnitude in metallicity, quantifying the relative dust abundance both through the direct line-of-sight visual extinction $A_V$ and the derived depletion level. We use a novel method to derive the DTG and DTM {\em mass} ratios for each GRB sightline, summing up the mass of all the depleted elements in the dust-phase. We find that the DTG and DTM mass ratios are both strongly correlated with the gas-phase metallicity and show a mild evolution with redshift as well. While these results are subject to a variety of caveats related to the physical environments and the narrow pencil-beam sightlines through the ISM probed by the GRBs, they provide strong implications for studies of dust masses to infer the gas and metal content of high-redshift galaxies, and particularly demonstrate the large offset from the average Galactic value in the low-metallicity, high-redshift regime. 
}

\maketitle

\section{Introduction}     
\label{sec:intro}


The baryon cycle, which includes processes such as the infall of neutral, pristine gas onto galaxies and their subsequent chemical enrichment with dust and metals, is one of the fundamental drivers of galaxy formation and evolution \citep{Tinsley80,Dayal18,Maiolino19,Peroux20}. In particular, dust serves as a catalyst for the production of molecular hydrogen H$_2$ on the surfaces of its grains \citep{Hollenbach71,Black87}, an important prerequisite for star formation. The fraction of dust to the overall gas and metal abundances is governed by the most predominant dust production channels, in addition to the efficiency of grain growth in the interstellar medium (ISM) or potential supernovae dust destruction or condensation scenarios \citep{Draine03,Dunne03,Mattsson12,Dwek16,Schneider16,DeVis21}. 

Long-duration gamma-ray bursts (GRBs hereafter) offer unique insights into the dust and metal abundances of the ISM in their star-forming host galaxies \citep{Savaglio03,Jakobsson04,Fynbo06,Prochaska07}. 
Since most GRBs are associated with the death of massive stars \citep[e.g.,][]{Woosley06}, they are linked to active star formation and thereby provide a reliable probe of star-forming galaxies through most of cosmic time \citep{Jakobsson06a,Kistler09,Robertson12,Tanvir12,Greiner15,Perley16,Ghirlanda22}. Moreover, GRBs are some of the most energetic, brightest cosmological sources known \citep{Gehrels09,Malesani23}, enabling detailed studies of the ISM in their host galaxies based on absorption-line spectroscopy, even out to $z\gtrsim 6$ \citep{Kawai06,Hartoog15,Saccardi23}. While recent observations of nearby GRBs connected to dynamical merger origins challenge this picture \citep{Rastinejad22,Levan23} and current evidence seem to point to a potential ``metallicity-bias'' limiting the production of GRBs in metal-rich environments at $z\lesssim 2$ \citep{Levesque10,Japelj16,Palmerio19,Graham19,Bjornsson19}, these effects are arguably small in the high-redshift universe.

Due to their physical origin, most GRBs trace the central, dense star-forming regions of their host galaxies, showing high neutral hydrogen (\hi) column densities of $N_{\rm HI}\geq 10^{20.3}$\,cm$^{-2}$ \citep{Vreeswijk04,Jakobsson06b,Fynbo09,Tanvir19,Selsing19}, known as damped Lyman-$\alpha$ absorbers (DLAs) \citep{Wolfe05}. Further, they enable studies of the molecular gas-phase \citep{Prochaska09,Kruhler13,Friis15,Bolmer19,Heintz19a,Heintz19c} and the dust-rich environments \citep{Zafar12,Fynbo14,Heintz17,Heintz19b,ZafarMoller19} of their host galaxies, which are much less frequently probed with absorbers in quasar sightlines \citep[though do appear more common in dusty, gas- or molecular-rich sightlines][Krogager et al. in prep.]{Heintz18,Ranjan20}. This is due to the typical higher impact parameters of quasar absorbers \citep{Peroux11,Krogager12,Christensen14,Rahmani16,Krogager17,Rhodin18}, which mostly probe the extended neutral gas reservoirs rather than the star-forming ISM \citep{Neeleman19,Heintz21}. Using GRBs as probes thus provides valuable insights into the dust and chemical abundances, in particular the dust-to-gas (DTG) and dust-to-metal (DTM) ratios, of the star-forming ISM of high-redshift galaxies.

Here we present new measurements and comprehensive analyses of the metal abundance and dust content of three GRB systems at $z>2$, studied through absorption-line spectroscopy of their bright optical/near-infrared afterglow. To complement these measurements, we further compile all GRB afterglows at $z\gtrsim 2$ observed with intermediate to high-resolution spectrographs for which similar measurements can be obtained, to provide the most comprehensive study to-date of the metallicity and dust content of GRB-selected star-forming galaxies through cosmic time. 

The paper is structured as follows. In Sect.~\ref{sec:sample} we present the observations of the three new GRBs, and describe the overall sample compilation. In Sect.~\ref{sec:analysis}, we detail the derivation of the metal abundances, the visual extinction, the dust-corrected metallicities and the DTG and DTM ratios for each GRB. In Sect.~\ref{sec:res} we present our results, and quantify the evolutionary trends of these properties with redshift and metallicity. Finally in Sect.~\ref{sec:conc} we discuss and conclude on our work, with particular emphasis on the implications of our results for galaxy evolution studies at high-redshift. 

Throughout the paper we assume the concordance $\Lambda$CDM cosmological model with $\Omega_{\rm m} = 0.315$, $\Omega_{\Lambda} = 0.685$, and $H_0 = 67.4$\,km\,s$^{-1}$\,Mpc$^{-1}$ \citep{Planck20}. We derive relative abundances of specific elements X and Y using the solar abundances as reference, [X/Y] = $\log (N_X/N_Y) - \log (N_X/N_Y)_\odot$, assuming the solar chemical abundances from \citet{Asplund21} based on the recommendations by \citet{Lodders09}. Unless indicated otherwise, all uncertainties are given at 1$\sigma$ confidence throughout the paper.

\section{Observations and sample compilation}     
\label{sec:sample}

In this work, we present measurements of the metal abundance and dust content in the sightlines of three new GRBs, GRBs\,190106A, 190919B, and 191011A, observed as part of the ESO-VLT STARGATE ToO Programme (PI: N. R. Tanvir). Further, we compile all GRB afterglows to-date, matching a few predefined criteria as detailed below. The large majority of the GRB afterglows in this work have been observed with the ESO VLT/X-shooter spectrograph \citep{Vernet11} as part of the XS-GRB survey programme \citep[PI: J. Fynbo;][]{Selsing19,Bolmer19}. Our compiled sample also includes eight bursts observed with the higher-resolution spectrographs UVES on the VLT \citep[6 out of 8;][]{Dekker00} and ESI on Keck \citep[2 out 8;][]{Sheinis02}. 
For this work, we require that the GRB afterglow have been observed with intermediate ($\mathcal{R} = 7000$) to high-resolution ($\mathcal{R} > 40,000$) spectrographs to ensure the robustness of the metal abundance measurements.
We additionally impose that at least the wavelength regions of the redshifted transitions of the refractory element \feii\ and the volatile elements \sii\ or \znii\ are covered and that the signal-to-noise (S/N) ratio is ${\rm S/N} > 3$ per resolution element in the regions surrounding these transitions. This is to further optimize the column density measurements of these transitions, and to ensure that we cover at least one heavily depleted and one volatile element to compute the dust-corrected gas-phase metallicities and the dust depletions in GRB-selected galaxies.   

\subsection{ESO-VLT/X-shooter observations}

GRBs\,190106A and 191011A were initially detected with the {\em Neil Gehrels Swift Observatory} \citep[{\em Swift} hereafter;][]{Gehrels04}, as reported by \citet{Sonbas19} and \citet{Laha19}, respectively. GRB\,190919B was detected with {\em INTEGRAL} \citep{Winkler03} as reported by \citet{Mereghetti19}. Following the detection of the optical counterpart, we obtained ultraviolet to near-infrared (300-2500 nm) spectroscopy of the GRB afterglows with the X-shooter spectrograph \citep{Vernet11} mounted on the European Southern Observatory (ESO) Very Large Telescope (VLT) Unit Telescope 2 (in 2019). The observations were carried out 11 hr (GRB\,190106A), 4.87 hr (GRB\,190919B), and 23.3 min (GRB\,191011A, using the rapid-response mode) after trigger. Each observation covered the ultraviolet to near-infrared simultaneously using the UVB, VIS and NIR arms of the VLT/X-shooter with slit-widths of $1\farcs 0$ (UVB) and $0\farcs 9$ (VIS,NIR) and nominal spectral resolutions of $\mathcal{R} = \lambda / \Delta \lambda = 5400$ (UVB), 8900 (VIS), and 5600 (NIR). The delivered spectral resolution are in most observations superior to the nominal, since the seeing full-width-at-half-maximum (FWHM) is considerably smaller than the slit width \citep{Selsing19}. 

The spectroscopic data were reduced and processed following a similar approach as described in \citet{Selsing19}. We use the version v. 3.5.3 of the ESO X-shooter pipeline \citep{Modigliani10}. The flux-calibrated 1D spectra were moved to the vacuum-heliocentric system in the process and corrected for Galactic extinction along the line-of-sight using the values from \citet{Schlafly11}.

\subsection{Sample compilation}

In addition to these bursts, we compile all GRB afterglow measurements from the literature following our criteria outlined above. This includes measurements from the pre-X-shooter era of the GRBs\,000926, 030226, 050730, 050820A, 050922C, 071031, 080413A, and 081008 \citep{Savaglio03,Shin06,Prochaska07,Piranomonte08,Ledoux09,DElia11,Wiseman17,ZafarMoller19}. Further, we consider all the GRBs observed as part of the XS-GRB legacy survey. Specifically, we adopt the column density and metallicity measurements from \cite{Bolmer19}, which includes GRB\,090809A up to GRB\,170202A. Beyond this, the GRB afterglows in our sample were all observed with the VLT/X-shooter as part of the STARGATE programme (PI: N. R. Tanvir). In addition to the three bursts presented above, we further include GRBs\,181020A, 190114A, and 210905A. GRBs\,181020A and 190114A have already been presented in \citet{Heintz19c}, but we here rederive their basic properties for consistency and homogeneity with the rest of the sample, and additionally include the new metallicity measurements for GRB\,210905A from \citet{Saccardi23}.  
We note that GRB\,180325A has been observed with X-shooter as part of the STARGATE programme as well \citep{Zafar18bump}. However, since the relevant metal line transitions are all heavily saturated, hindering robust metallicity and depletion measurements, we exclude this burst from the sample. The full GRB afterglow sample is comprised of 36 bursts, with their physical properties summarized in Table~\ref{tab:results}.

\begin{table*}
\caption{Overview of the absorption-derived GRB host galaxy ISM properties.}              
\label{tab:results}      
\begin{tabular}{l c c c c c c c c c}
\hline\vspace{0.1cm}
GRB & $z_{\rm GRB}$ & $A_V$ (SED) & $\log N_{\rm HI}$ & [X/H] & X & [Zn/Fe] & [M/H]$_{\rm tot}$  & Telescope/ & Ref. \\  
 & & (mag) & (cm$^{-2}$) &&&&& Instrument & \\
 (1) &  (2) &  (3) &  (4) &  (5) &  (6) &  (7) &  (8) &  (9) &  (10) \\
\hline    
000926  &  	 2.0380  & 	  0.38 $\pm$ 0.05  &  	  21.30 $\pm$ 0.20   &  	  -0.11 $\pm$ 0.21  &  	 Zn  & 	 1.06 $\pm$ 0.18  &  	 0.20 $\pm$ 0.28  & 	  Keck/ESI  & 	 (1,2) \\
 	 030226  &  	 1.9870  & 	  --  &  	  20.50 $\pm$ 0.30   &  	  -0.94 $\pm$ 0.30  &  	 Si  & 	 -0.18 $\pm$ 0.12  &  	 -1.07 $\pm$ 0.31  & 	  Keck/ESI  & (3,4) \\
 	 050730  &  	 3.9690  & 	  0.12 $\pm$ 0.02  &  	  22.10 $\pm$ 0.10   &  	  -2.18 $\pm$ 0.11  &  	 S  & 	 0.08 $\pm$ 0.05  &  	 -2.31 $\pm$ 0.18  & 	  VLT/UVES  &  (2,5,6) \\
 	 050820A  &  	 2.6150  & 	  0.27 $\pm$ 0.04  &  	  21.05 $\pm$ 0.10   &  	  -0.39 $\pm$ 0.10  &  	 Zn  & 	 0.83 $\pm$ 0.05  &  	 -0.49 $\pm$ 0.10  & 	  VLT/UVES+HIRES  & 	 (2,4,5,6) \\
 	 050922C  &  	 2.1990  & 	  0.09 $\pm$ 0.03  &  	  21.55 $\pm$ 0.10   &  	  -2.09 $\pm$ 0.12  &  	 Si  & 	 0.18 $\pm$ 0.46  &  	 -1.92 $\pm$ 0.26  & 	  VLT/UVES  & 	 (5,7) \\
 	 071031  &  	 2.6920  & 	  $<0.07$  &  	  22.15 $\pm$ 0.05   &  	  -1.76 $\pm$ 0.05  &  	 Zn  & 	 0.04 $\pm$ 0.03  &  	 -1.75 $\pm$ 0.09   & 	  VLT/UVES  & 	 (5) \\
 	 080413A  &  	 2.4330  & 	  $<0.59$ &  	  21.85 $\pm$ 0.15   &  	  -1.63 $\pm$ 0.16  &  	 Zn  & 	 0.13 $\pm$ 0.07  &  	 -1.60 $\pm$ 0.18  & 	  VLT/UVES  & 	 (5) \\
 	 081008  &  	 1.9685  & 	  $<0.08$   &  	  21.11 $\pm$ 0.10   &  	  -0.52 $\pm$ 0.11  &  	 Zn  & 	 0.55 $\pm$ 0.04  &  	 -0.51 $\pm$ 0.17  & 	  VLT/UVES  & 	 (6,8) \\
 	 090809A  &  	 2.7373  & 	  0.11 $\pm$ 0.04  &  	  21.48 $\pm$ 0.07   &  	  -0.86 $\pm$ 0.13  &  	 Zn  & 	 0.75 $\pm$ 0.21  &  	 -0.46 $\pm$ 0.15   & 	  VLT/X-shooter  & 	 (9) \\
 	 090926A  &  	 2.1069  & 	  $<0.03$  &  	  21.58 $\pm$ 0.01   &  	  -1.97 $\pm$ 0.11  &  	 Zn  & 	 0.84 $\pm$ 0.05  &  	 -1.72 $\pm$ 0.05 & VLT/X-shooter  & 	 (9,10) \\
 	 100219A  &  	 4.6676  & 	  0.15 $\pm$ 0.05  &  	  21.28 $\pm$ 0.02   &  	  -1.24 $\pm$ 0.05  &  	 S  & 	 0.12 $\pm$ 0.30  &  	 -1.16 $\pm$ 0.11   & 	 VLT/X-shooter  & 	 (9,11) \\
 	 111008A  &  	 4.9910  & 	  0.13 $\pm$ 0.05  &  	  22.39 $\pm$ 0.01   &  	  -1.48 $\pm$ 0.31  &  	 Zn  & 	 0.22 $\pm$ 0.10  &  	 -1.79 $\pm$ 0.10 & 	 VLT/X-shooter  & 	 (9,12) \\
 	 111107A  &  	 2.8930  & 	  $<0.15$  &  	  21.10 $\pm$ 0.04   &  	  -0.74 $\pm$ 0.2  &  	 Si  & 	 0.70 $\pm$ 0.55  &  	 -0.28 $\pm$ 0.45  & 	VLT/X-shooter  & 	 (9) \\
 	 120119A  &  	 1.7285  & 	  1.06 $\pm$ 0.02  &  	  22.44 $\pm$ 0.12   &  	  -1.03 $\pm$ 0.25  &  	 Zn  & 	 0.93 $\pm$ 0.24  &  	 -0.79 $\pm$ 0.42  & VLT/X-shooter  & 	 (6) \\
 	 120327A  &  	 2.8143  & 	  0.05 $\pm$ 0.02  &  	  22.07 $\pm$ 0.01   &  	  -1.49 $\pm$ 0.04  &  	 Zn  & 	 0.34 $\pm$ 0.03  &  	 -1.34 $\pm$ 0.02  & 	VLT/X-shooter  & 	 (9,13) \\
 	 120716A  &  	 2.4874  & 	  0.30 $\pm$ 0.15  &  	  21.73 $\pm$ 0.03   &  	  -0.71 $\pm$ 0.06  &  	 Zn  & 	 0.60 $\pm$ 0.15  &  	 -0.57 $\pm$ 0.08 & 	VLT/X-shooter  & 	 (9) \\
 	 120815A  &  	 2.3582  & 	  0.19 $\pm$ 0.04  &  	  22.09 $\pm$ 0.01   &  	  -1.45 $\pm$ 0.03  &  	 Zn  & 	 0.95 $\pm$ 0.04  &  	 -1.23 $\pm$ 0.03  & 	VLT/X-shooter  &  (9,14) \\
 	 120909A  &  	 3.9290  & 	  0.16 $\pm$ 0.04  &  	  21.82 $\pm$ 0.02   &  	  -1.06 $\pm$ 0.12  &  	 S  & 	 1.15 $\pm$ 0.09  &  	 -0.29 $\pm$ 0.10  & 	VLT/X-shooter  & 	 (9) \\
 	 121024A  &  	 2.3005  & 	  0.26 $\pm$ 0.07  &  	  21.78 $\pm$ 0.02   &  	  -0.76 $\pm$ 0.06  &  	 Zn  & 	 0.73 $\pm$ 0.07  &  	 -0.68 $\pm$ 0.07 & 	VLT/X-shooter  & 	 (9,15) \\
 	 130408A  &  	 3.7579  & 	  0.12 $\pm$ 0.03  &  	  21.90 $\pm$ 0.01   &  	  -1.48 $\pm$ 0.07  &  	 Zn  & 	 0.29 $\pm$ 0.07  &  	 -1.46 $\pm$ 0.05   & 	  VLT/X-shooter  & (9) \\
 	 130606A  &  	 5.9127  & 	  $<0.03$  &  	  19.88 $\pm$ 0.01   &  	  -1.83 $\pm$ 0.10  &  	 Si  & 	 0.49 $\pm$ 0.10  &  	 -1.58 $\pm$ 0.08   & 	  VLT/X-shooter  & 	(9,16) \\
 	 140311A  &  	 4.9550  & 	  0.07 $\pm$ 0.03  &  	  22.30 $\pm$ 0.02   &  	  -1.65 $\pm$ 0.14  &  	 Zn  & 	 0.23 $\pm$ 0.11  &  	 -2.00 $\pm$ 0.11   & 	  VLT/X-shooter  & 	(9) \\
 	 141028A  &  	 2.3333  & 	  0.13 $\pm$ 0.09  &  	  20.39 $\pm$ 0.03   &  	  -1.64 $\pm$ 0.13  &  	 Si  & 	 -0.04 $\pm$ 0.26  &  	 -1.62 $\pm$ 0.28  & 	  VLT/X-shooter  & 	 (9) \\
 	 141109A  &  	 2.9940  & 	  0.16 $\pm$ 0.04  &  	  22.18 $\pm$ 0.02   &  	  -1.63 $\pm$ 0.06  &  	 Zn  & 	 0.61 $\pm$ 0.05  &  	 -1.37 $\pm$ 0.05   & 	  VLT/X-shooter  & 	(9) \\
 	 150403A  &  	 2.0571  & 	  0.12 $\pm$ 0.02  &  	  21.73 $\pm$ 0.02   &  	  -1.04 $\pm$ 0.04  &  	 Zn  & 	 0.47 $\pm$ 0.05  &  	 -0.92 $\pm$ 0.05 & 	  VLT/X-shooter  & 	 (9) \\
 	 151021A  &  	 2.3297  & 	  0.20 $\pm$ 0.03  &  	  22.14 $\pm$ 0.03   &  	  -0.98 $\pm$ 0.07  &  	 Zn  & 	 0.69 $\pm$ 0.07  &  	 -0.97 $\pm$ 0.07  & 	  VLT/X-shooter  & 	(9)\\
 	 151027B  &  	 4.0650  & 	  0.10 $\pm$ 0.05  &  	  20.54 $\pm$ 0.07   &  	  -0.76 $\pm$ 0.17  &  	 S  & 	 0.49 $\pm$ 0.64  &  	 -0.59 $\pm$ 0.27  & 	  VLT/X-shooter  & 	(9) \\
 	 160203A  &  	 3.5187  & 	  $<0.10$  &  	  21.74 $\pm$ 0.02   &  	  -1.31 $\pm$ 0.04  &  	 S  & 	 0.37 $\pm$ 0.18  &  	 -0.92 $\pm$ 0.04  & 	  VLT/X-shooter  & 	(9, 17) \\
 	 161023A  &  	 2.7100  & 	  0.09 $\pm$ 0.03  &  	  20.95 $\pm$ 0.01   &  	  -1.23 $\pm$ 0.03  &  	 S  & 	 0.44 $\pm$ 0.04  &  	 -1.05 $\pm$ 0.04  & 	  VLT/X-shooter  & 	(9,18) \\
 	 170202A  &  	 3.6456  & 	  0.08 $\pm$ 0.03  &  	  21.53 $\pm$ 0.04   &  	  -1.28 $\pm$ 0.09  &  	 S  & 	 0.76 $\pm$ 0.23  &  	 -1.02 $\pm$ 0.13  & 	  VLT/X-shooter  & 	(9)\\
 	 181020A  &  	 2.9379  & 	  0.29 $\pm$ 0.02  &  	  22.24 $\pm$ 0.03   &  	  -1.57 $\pm$ 0.11  &  	 Zn  & 	 0.76 $\pm$ 0.14  &  	 -1.20 $\pm$ 0.08  & 	  VLT/X-shooter  & 	(19,20) \\
 	 190106A  &  	 1.8599  & 	  0.27 $\pm$ 0.03  &  	  21.00 $\pm$ 0.04   &  	  -0.33 $\pm$ 0.10  &  	 Zn  & 	 1.13 $\pm$ 0.10  &  	 -0.40 $\pm$ 0.10  & 	  VLT/X-shooter  & 	 (20) \\
    	 190114A  &  	 3.3764  & 	  0.34 $\pm$ 0.01  &  	  22.19 $\pm$ 0.05   &  	  -1.44 $\pm$ 0.24  &  	 S  & 	 1.06 $\pm$ 0.08  &  	 -1.17 $\pm$ 0.06  & 	  VLT/X-shooter  & 	(19,20) \\
 	 190919B  &  	 3.2241  & 	  $<0.03$  &  	  21.49 $\pm$ 0.03   &  	  -1.40 $\pm$ 0.16  &  	 S  & 	 0.33 $\pm$ 0.33  &  	 -1.25 $\pm$ 0.15  & 	  VLT/X-shooter  & 	 (20) \\
 	 191011A  &  	 1.7204  & 	  0.43 $\pm$ 0.03  &  	  21.65 $\pm$ 0.08   &  	  -0.95 $\pm$ 0.11  &  	 Zn  & 	 0.33 $\pm$ 0.09  &  	 -0.63 $\pm$ 0.08   & 	  VLT/X-shooter  & 	 (20) \\
 	 210905A  &  	 6.3118  & 	  $<0.02$  &  	  21.10 $\pm$ 0.10   &  	  -1.71 $\pm$ 0.11  &  	 Si  & 	 0.33 $\pm$ 0.09  &  	 -1.72 $\pm$ 0.13  & VLT/X-shooter  & 	 (21) \\
\hline          
\end{tabular}
\textbf{Notes.} This Table is composed mainly of GRB afterglow measurements from the literature, in addition to the three new bursts analyzed here: GRBs\,190106A, 190919B, and 191011A. Col.~(1): GRB names. Col.~(2): Spectroscopic redshift of the absorption system. Col.~(3): Visual extinction derived from the SED. Upper limits are reported at $1\sigma$. Col.~(4): \hi\ column density derived from the Lyman-$\alpha$ absorption feature. Col.~(5): Metallicity of the element X. Col.~(6): Element X used for the metallicity and depletion measurements. Col.~(7): Zinc-over-iron depletion level. Col.~(8): Dust-corrected metallicity. Col.~(9): Telescope and instrument with which the GRB afterglows were observed. Col.~(10): References for the first afterglow spectra presentations and subsequent measurements adopted in this work.     
{\bf References.} (1)~\cite{Savaglio03}; (2)~\cite{ZafarMoller19}; (3)~\cite{Shin06}; (4)~\cite{Prochaska07}; (5)~\cite{Ledoux09}; (6)~\cite{Wiseman17}, (7)~\cite{Piranomonte08}; (8)~\cite{DElia14}; (9)~\cite{Bolmer19}; (10)~\cite{DElia10}; (11)~\cite{Thoene13}; (12)~\cite{Sparre14}; (13)~\cite{DElia14}; (14)~\cite{Kruhler13}; (15)~\cite{Friis15}; (16)~\cite{Hartoog15}; (17)~Pugliese et al. (submitted); (18)~\cite{deUgarte18}; (19)~\cite{Heintz19c}; (20)~This work; (21)~\cite{Saccardi23}. 

\end{table*}

\section{Methods and Analysis} \label{sec:analysis}

\subsection{Metal abundances} \label{ssec:abslines}

To model the absorption lines of the three new GRBs\,190106A, 190919B, and 191011A considered here, we use the Python module \texttt{VoigtFit} \citep{Krogager18}. This code takes the observed spectra as input, convolves the Voigt-profiles to match the delivered spectral resolution, and provides the best-fit column density $N$ and broadening parameter $b$ for each transition separately for each of the identified velocity components. We model and tie $b$ and the velocity structure for all the low-ionization transitions, based on the assumption that they physically trace the bulk of the neutral gas \citep[e.g.,][]{Prochaska97}, that is $N_{\rm FeII} = N_{\rm Fe}$. This is physically motivated since the ionization potentials of the neutral ions considered here are below that of hydrogen (13.6 eV), such that they will predominantly be in the singly-ionized state in the neutral gas-phase. Further, these absorption-line abundances have been found to not be influenced by photoionization from the GRB prompt emission since they typically probe gas in the ISM on kpc scales away from the GRB progenitor \citep{Vreeswijk07,Prochaska07,Prochaska08,Ledoux09,Heintz18}. 

The absorption-line spectra and the best-fit models are shown in Appendix A. To determine the gas-phase metallicity, ${\rm [X/H]} = \log (N_{\rm X}/N_{\rm H}) - \log (N_{\rm X}/N_{\rm H})_{\odot}$, for each burst, we first fit the \hi\ column density based on the broad damped Lyman-$\alpha$ absorption trough. Then, we rely primarily on the volatile elements Zn or S to determine the metal abundances. To determine the overall dust depletion level, quantified via ${\rm [Zn/Fe]} = \log (N_{\rm Zn}/N_{\rm Fe}) - \log (N_{\rm Zn}/N_{\rm Fe})_{\odot}$ \citep[see e.g.]{DeCia18}, we either derive it directly from the measured abundances or, in the case that Zn is inaccessible, we determine the expected [Zn/Fe]$_{\rm exp}$ following the relations from \citet{DeCia18} as described below. The derived \hi\ column densities, [X/H] and [Zn/Fe] for each of the bursts in the full sample are summarized in Table~\ref{tab:results}.

\subsection{Dust-corrected metallicities} \label{ssec:mhtot}

Due to mild or strong dust depletion of volatile and refractory elements, respectively, a significant fraction of the metals will be missing from the observed gas-phase abundances, $[\text{X/H}]_{\text{obs}}$. To gauge the actual metal abundance of the GRB host galaxies we therefore need to take into account the metals both in the dust- and gas-phase. The model for the expected relative abundances can be expressed as
\begin{equation}
    {\rm [X/H]_{exp}} = \delta_X + {\rm [M/H]_{tot}} = A2_X + B2_X \times {\rm [Zn/Fe]} + {\rm [M/H]_{tot}} \label{eq:abundance}
\end{equation}
where $\delta_{\rm X}$ is the dust depletion of element X \citep[see e.g.,][]{DeCia16,Konstantopoulou22} and $A2_X$ and $B2_X$ are empirically computed linear depletion parameters, here taken from Konstantopoulou et al. (subm., see also \citealt{DeCia16}). For all cases, $\delta_{\rm X} \leq 0$, with more negative values indicating higher depletion levels. For each source, we can thus derive the total, dust-corrected metallicity [M/H]$_{\rm tot}$ and the overall strength of dust depletion, [Zn/Fe]$_{\rm fit}$, by performing a fit minimizing the difference between the observed relative abundances $[\text{X/H}]_{\text{obs}}$ and the relative abundances $[\text{X/H}]_{\text{exp}}$ given by Eq. \eqref{eq:abundance}. 

To sample the posterior distribution of the best fit parameters, we use the implementation of a Dynamical Nested Sampling algorithm provided by the \texttt{dynesty} package \citep{skilling04, dns18, dynesty20}. This type of sampling algorithm has the benefits of focused Bayesian posterior estimation as performed by Markov Chain Monte Carlo (MCMC) samplers while retaining the ability to determine marginal likelihoods for model comparison like other Nested Sampling algorithms. We use a uniform prior between $0<$ [Zn/Fe] $<1.7$ for the depletion strength parameter, where the upper limit is motivated by the strongest levels of depletion in Galactic sightlines presented by \citet{Jenkins09}. For the dust-corrected metallicity, we also use a uniform prior, but allow it to run from $\text{[M/H]}_{\rm tot}=-3.0$ to $1.0$ which encompasses all known Milky Way sightlines and high-$z$ GRB absorption systems. As expected, the dust-corrected metallicities are overall higher than the metallicities inferred using Zn, S or Si as tracers. For GRBs\,190106A, 190919B, and 191011A, we derive dust-corrected gas-phase metallicities of [M/H]$_{\rm tot} = -0.40\pm 0.10$, $-1.25\pm 0.15$, and $-0.63\pm 0.08$, respectively. The dust-corrected metallicities reported in \citet{Bolmer19} were computed following a similar approach, and the pre-X-shooter GRB sample were re-analysed to compute dust-corrected metallicities following \citet{DeCia18}. The full sample covers a large range in metallicities of $\text{[M/H]}_{\rm tot}=-2.3$ to $0.2$, i.e. $0.5\%$ to $150\%$ solar abundances.

\subsection{Line-of-sight visual extinction} \label{ssec:avext}

To determine the total integrated amount of dust in the GRB host-galaxy sightline, we model the extinction of the observed afterglow spectral energy distribution (SED) \citep[e.g.,][]{Watson06,Schady10,Zafar11,Greiner11,Covino13,Zafar18ext}. Since the optical afterglows of GRBs follow an underlying smooth, temporally-varying power-law \citep{Sari98}, it is possible to very accurately measure the visual extinction $A_V$ and the total-to-selective extinction $R_V$. In contrast, the dust in DLAs in quasar sightlines are more difficult to disentangle due to the potential additional extinction of the background quasar spectrum and the uncertainties in functional shape. Following \citet{Heintz19a}, we adopt from the {\em Swift}/XRT repository\footnote{\url{https://www.swift.ac.uk/xrt_spectra/}} the X-ray spectral slope in photon units, $\Gamma$, as derived from the {\it Swift}/XRT afterglow spectrum as prior for the intrinsic spectral slope converted to a function of wavelength as $F_{\lambda} = F_{0}\lambda^{\Gamma - \Delta \beta - 3}$ and allow for the synchrotron cooling spectral break $\Delta \beta$ to take a value of $\Delta \beta = 0.0$ or 0.5 \citep{Sari98}.

We model the observed, dust-extinguished afterglow as $F^{\rm obs}_\lambda = F_\lambda \times 10^{-0.4 A_\lambda}$ where $A_\lambda$ is the extinction as a function of wavelength. To determine the visual extinction $A_V$ for three new GRBs, we assume the average SMC extinction law \citep[as parametrized by][]{Gordon03} due to the lack of any evidence for the rare 2175\,\AA\ extinction bump or an unusual steep (or flat) reddening curve in this sample (see also Appendix~\ref{sec:app}). This is also in line with past GRB observations \citep{Savaglio04,Perley08,Kann06,Kann10,Friis15,Zafar18ext,Corre18}, where the 2175\,\AA\ dust bump is only observed in a handful of cases \citep{Zafar12,Zafar18bump,Heintz19b}. More exotic extinction curves have also been seen, either being extremely steep as in the case of GRB\,140506A \citep{Fynbo14,Heintz17} or in a few bursts showing more flat, ``grey'' dust distributions \citep[e.g.,][]{Stratta04,Stratta05,Perley08}. However, in the majority of cases, an SMC-like extinction curves appear to be the most prevalent considering GRBs with spectral coverage from X-rays to the ultraviolet and near-infrared \citep{Zafar18ext}.

We normalize the intrinsic afterglow spectrum to the flux level in the NIR arm around the wavelength region of the typical $K$-band ($\sim 2\mu$m). We fix the redshift to $z_{\rm GRB}$ and thereby only fit for $A_V$ for each case. We derive $A_V = 0.83\pm 0.03$, $<0.03$ ($3\sigma$), and $0.48\pm 0.13$\,mag for GRBs\,190106A, 190919B, and 191011A, respectively. The spectra and best-fit extinction curve models are shown in Appendix A and the results are summarized for the full sample in Table~\ref{tab:results}.

\section{Results and interpretations}     
\label{sec:res}

\subsection{Dust-corrected metallicity evolution with redshift}

\begin{figure*}[!t]
    \centering
    \includegraphics[width=0.8\textwidth]{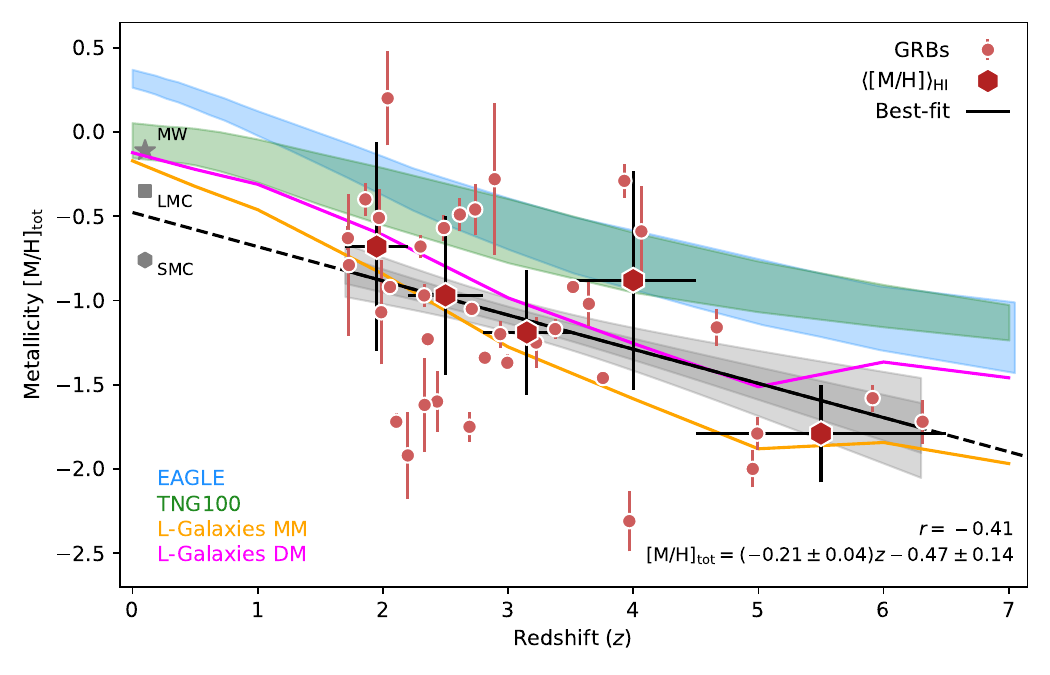}
    \caption{Dust-corrected metallicity [M/H]$_{\rm tot}$ as a function redshift for the GRB-selected galaxies. The small red data points show individual measurements, and the large red hexagons represent the \hi-weighted means with redshift where the errorbars denote the redshift interval and $1\sigma$ dispersion, respectively. The best-fit relation ${\rm [M/H]}_{\rm tot}(z) = (-0.21\pm 0.04)z - (0.47\pm 0.14)$ is shown as the solid black line, with the dark- and light-shaded gray regions indicating the 1 and $2\sigma$ confidence intervals. For comparison, we overplot the average dust-corrected metallicities of MW, LMC, and SMC sightlines and predictions from the compiled set of simulations from \citet{Yates21b}: the ``default model'' (DM) and the ``modified model (MM)'' from the L-Galaxies simulations \citep{Henriques20,Yates21a}, the EAGLE simulations \citep{Crain15,Schaye15}, and the IllustrisTNG-100 \citep{Pillepich18,Springel18}. Generally, all observations and simulations seem to find similar slopes of $\Delta\log{\rm (O/H)}/\Delta z \approx 0.1-0.3$. However, only the L-Galaxies simulations are able to reproduce the lower average metallicities inferred from the GRB sightlines.}
    \label{fig:metz}
\end{figure*}

In Fig.~\ref{fig:metz} we present the redshift evolution of the dust-corrected metallicities measured for the 36 GRBs in our sample, spanning $z_{\rm GRB} = 1.7 - 6.3$. We perform a linear fit of the data, including the errors on [M/H]$_{\rm tot}$, from which we find ${\rm [M/H]}_{\rm tot}(z) = (-0.21\pm 0.04)z - (0.47\pm 0.14)$. This slope is slightly steeper and the intercept at $z=0$ higher than inferred previously for GRBs, and implies a significant evolution compared to previous results \citep{Cucchiara15}. Their work did not consider the dust-corrected metallicities, however, which would explain the offset in the intercept. The evolution of [M/H]$_{\rm tot}$ with redshift is observed to be even steeper for DLAs in quasar sightlines \citep{DeCia16,DeCia18}, with a slope of $-0.32\pm 0.04$ as inferred from the dust-corrected metallicities of their large sample. The observed lower intercept of the quasar-DLA relation \citep{DeCia18} is also expected since GRB-selected samples are weighted towards more metal-rich galaxies compared to quasar DLAs due to their lower impact parameters \citep{Fynbo08,Arabsalmani15}. We caution that our inferred metallicity evolution with redshift is largely driven by the high-redshift points at $z\gtrsim 5$, which is still sparsely populated and may be subject to a more severe selection bias as they will appear optically ``dark'' \citep[e.g.,][]{Fynbo01}. 

To account for the low-metallicity GRB systems which carry less gas in our analysis, we derive the average \hi-weighted metallicities $\langle {\rm [M/H]}\rangle_{\rm HI}$, defined as
\begin{equation}
    \langle {\rm [M/H]}\rangle_{\rm HI} = \frac{\sum (10^{\rm [M/H]_{tot}} \times N_{\rm HI})}{\sum N_{\rm HI}}
\end{equation}
and shown in Fig.~\ref{fig:metz}. The errorbars on $z$ represent the span on the redshift range and the standard deviation on $\langle {\rm [M/H]}\rangle_{\rm HI}$. We divide the sample into larger redshift bins at higher redshifts, considering points at $z=1.7-2.2$, $z=2.2-2.8$, $z=2.8-3.5$, $z=3.5-4.5$, and $z=4.5-6.3$, respectively, to account for the sparser number of sources in our sample at early cosmic epochs.

To put our results into context, we compare our measurements to recent simulations mapping the chemical enrichment and the metal mass density in galaxies across cosmic time. In particular, we adopt the ``default model'' (DM) and the ``modified model (MM)'' from the L-Galaxies simulations \citep{Henriques20,Yates21a}, the EAGLE simulations \citep{Crain15,Schaye15}, and the IllustrisTNG-100 \citep{Pillepich18,Springel18}, as compiled and described in detail by \citet{Yates21b}. While all simulations seem to find slopes for the redshift evolution of ${\rm [M/H]}_{\rm tot}(z)$ in agreement with our measurements, the EAGLE and TNG-100 simulations return a higher normalization, in particular at $z\gtrsim 5$. This could be due to an over-production or over-retention of metals inside galaxies in these particular simulations, which could indicate that EAGLE and TNG-100 contain an overabundance of massive galaxies. Overall, the L-Galaxies MM galaxy evolution models seem to best reproduce the data, with near-solar metallicity at $z\sim 0$ and reaching ${\rm [M/H]}_{\rm tot}=-2$ at $z\gtrsim 5$. These lower cosmic metallicities are achieved in L-Galaxies MM through highly efficient removal of metal-rich material from galaxies by supernova-driven galactic winds (see \citealt{Yates21a}). We also note the particular metal-poor GRB system, GRB\,050730 with ${\rm [M/H]}_{\rm tot} = -2.31\pm 0.18$ at $z=3.969$ which neither of the models are able to reproduce and is also substantially offset from the underlying metallicity-evolution probed by the GRB sample. This is most likely related to the selection effects of the high-resolution VLT/UVES sample \citep{Ledoux09}, but overall still imply that very metal-poor galaxies exist at $z=4$. These observations thus provide new statistics on galaxy properties and their population scatter, which has to be considered in most recent simulation frameworks.

Further, while GRBs do not have the same biases as emission-selected galaxies and thus provide a more complete census of star-forming galaxies at high-$z$ \citep{Fynbo08}, they may show an aversion to massive, metal-rich host galaxies at $z<2$ \citep{Perley13,Schulze15,Vergani15,Vergani17,Japelj16,Palmerio19}. This would explain the lower intercept at $z=0$ of the GRB absorbers compared to simulations. However, this does not explain the offset at higher redshifts ($z>3$), where GRBs are found to robustly trace the star-forming galaxy population.
We also that note that we observe a substantial scatter in the dust-corrected metallicities for a given redshift in the GRB sample, which is not recovered by any of the simulations. This observed scatter potentially seems to decrease with increasing redshift, though this may simply be due to lack of statistics.

While GRBs provide unique measures of the gas-phase metallicities in the ISM of galaxies out to high redshifts, other recent efforts to characterize the metallicities of galaxies out to and beyond $z\approx 4$ have recently been carried out in emission as well \citep[e.g.,][]{Sanders20,Cullen21,Heintz22met,Curti23}. 
New approaches to derive metallicities based on FIR line features such as [O\,{\sc iii}]$-88\mu$m detectable by ALMA at the same epoch has also recently been established \citep[e.g.,][]{Jones20}. However, these galaxies are luminosity-selected, and therefore represent only the most massive and metal-rich population of star-forming galaxies at these redshifts. Indeed, \citet{Cullen21} find oxygen abundances in the range $12+\log{\rm (O/H)} = 7.7-8.4$ (i.e. ${\rm [M/H]} \approx{} -0.80$ to $0.0$) for galaxies at $z\approx 3$, which is systematically higher than the average GRB absorption-based metallicity at this redshift (likely related to their high stellar masses, $M_\star > 10^{8.5}\,M_\odot$). Similarly, \citet{Jones20} derive oxygen abundances in the range $12+\log{\rm (O/H)} = 7.5-8.2$ (i.e. ${\rm [M/H]} = -1.2$ to $-0.5$) for galaxies at $z\gtrsim 7$, representing only the top $15\%$ most metal-rich GRB host galaxies at these redshifts. Moreover, metallicity measurements from nebular emission lines, such as those taken by \citet{Cullen21}, are strongly dependent on the strong-line diagnostics used \citep{Kewley08}, and generally only represent the metals in H\,{\sc ii} regions, which may be a poor reflection of the abundances in the more diffuse ISM. Further comparing the redshift evolution of luminosity-selected galaxies, we find that the evolution inferred for GRBs is slightly steeper than the slope $\Delta\log{\rm (O/H)}/\Delta z \approx -0.11\pm 0.02$ measured by \citet{Sanders21} from the MOSDEF galaxy survey, the latter also consistent with the results of \citet{Jones20} from $z\approx 0-8$. This discrepancy (at $2\sigma$ confidence) can potentially be due to the different galaxy luminosity distributions and mass ranges probed with either approach or attributed to the different evolution of the stellar and gas-phase metallicities in galaxies \citep[e.g.,][]{Yates21b,Fraser-McKelvie22} or total integrated vs. line-of-sight effects \citep[e.g.,][]{Arabsalmani23}.

\subsection{The evolution of the dust-to-metals ratio}

\begin{figure}
    \centering
    \includegraphics[width=9cm]{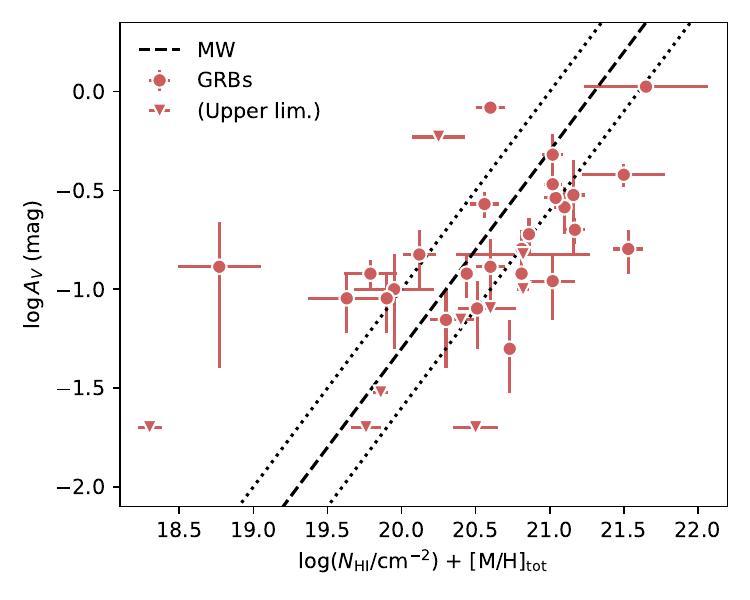}
    \caption{$A_V$ vs. the equivalent metal column density, $\log N_{\rm HI} + {\rm [M/H]_{tot}}$, i.e. the dust-to-metals (DTM) ratio. The red symbols show the GRB sample where the triangles denote $1\sigma$ upper limits. The dashed and dotted lines represent the average MW ratio and the scatter \citep{Watson11}. GRB sightlines probe a large range in DTM ratios, with an average around the Galactic mean value.}
    \label{fig:avnhimet}
\end{figure}

One way of inferring the dust-to-metal (DTM) ratio in the GRB sightlines is by using the direct measurements of $N_{\rm HI}$ and [M/H]$_{\rm tot}$ to trace the equivalent metal column density, $\log N_M = \log N_{\rm HI} + {\rm [M/H]_{tot}}$, and the visual extinction $A_V$, which traces the total integrated dust column in the line-of-sight. This is presented in Fig.~\ref{fig:avnhimet}. Overall, we observe a substantial scatter with respect to a constant DTM ratio (dashed curve), with an average value in the GRB sample of ${\rm DTM}_{\rm SED} = \log A_V - (\log N_{\rm HI} + {\rm [M/H]_{tot}}) = 4\times 10^{-22}$\,mag\,cm$^{2}$. This is consistent with previous GRB measurements \citep[e.g.,][]{Zafar13,Wiseman17,ZafarMoller19}, and slightly lower than the DTM measured for the Milky Way, ${\rm DTM}_{\rm Gal} = 4.5\times 10^{-22}$\,mag\,cm$^{2}$ \citep{Watson11}, though still consistent within the uncertainties. Notably, $A_V$ does not seem to decrease significantly below $\log N_{\rm HI} + {\rm [M/H]_{tot}} < 20.0$, which might suggest that a non-negligible fraction of the dust in the line-of-sight is not associated with the neutral gas-phase and instead might originate in the more ionized medium.

Further, we can infer the mass of the elements $X_i$ in the dust-phase relative to the total metal mass in the line-of-sight, the DTM {\em mass} ratio, based on the total, dust-corrected metallicity and depletion level inferred for each GRB host galaxy. Following the approach described in Konstantopoulou et al. (subm.) we derive
\begin{equation}\label{eq:dtm}
    {\rm DTM_{mass}} = \frac{M_{\rm dust}}{M_{\rm metals}} = \frac{\sum_{X_i}(1-10^{\delta_{X_i}})10^{\rm ([X_i/H]_\odot + [M/H]_{\rm tot})} W_{X_i}}{\sum_{X_i} 10^{\rm ([X_i/H]_\odot + [M/H]_{\rm tot})} W_{X_i}}
\end{equation}
where $\delta_{\rm X_i}$ is the dust depletion of each element X \citep[see also][]{DeCia16,Konstantopoulou22}, $W_{X_i}$ the atomic weight, and $10^{([X_i/H]_\odot + [M/H]_{\rm tot})}$ represents the total metal column of each element X. It is evident that the dust-corrected metallicity [M/H]$_{\rm tot}$ cancels out such that the DTM$_{\rm mass}$ ratio is independent of the overall metallicity of the system. While only a subset of all the expected metals in the dust- and gas-phase are measured, we calculate the total contribution from each element based on the overall depletion level [Zn/Fe]. We derive the depletion for each element X from the empirical relations, $\delta_X = A2_X + B2_X \times {\rm [Zn/Fe]}$, assuming the empirical depletion coefficients $A2_X$ and $B2_X$ from \citet{Konstantopoulou22}. The resulting DTM$_{\rm mass}$ ratios span $0.03\pm 0.01$ (GRB\,071031) to $0.41\pm 0.05$ (GRB\,190106A), generally lower than the Milky Way average of DTM = 0.45, as listed in Table~2.

\begin{figure}[!t]
    \centering
    \includegraphics[width=9.2cm]{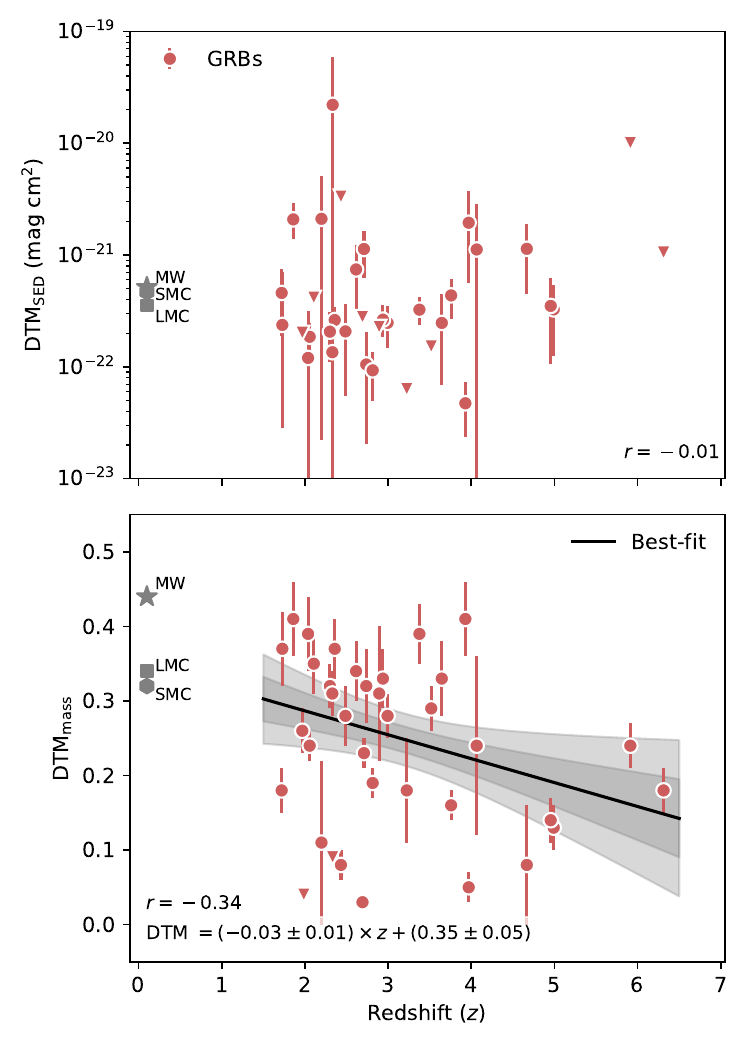}
    \caption{SED-derived dust-to-metals ratio, DTM$_{\rm SED} = \log A_V - (\log N_{\rm HI}+{\rm [M/H]_{\rm tot}})$ (top) and depletion-derived DTM$_{\rm mass}$ ratio (bottom) as a function of redshift. Red dots (measurements) and triangles ($1\sigma$ upper limits) denote the GRB host-galaxy absorbers. Grey symbols show the equivalent values for the MW, LMC and SMC. The Pearson correlation coefficients $r$ are marked for each data set. We observe no clear evolution of DTM$_{\rm SED}$ with redshift, but a mild evolution of DTM$_{\rm mass}$, with the best-fit relation shown as the solid black line, with the dark- and light-shaded gray regions indicating the 1 and $2\sigma$ confidence intervals. }
    \label{fig:DTMz}
\end{figure}

In Fig.~\ref{fig:DTMz} we show the evolution of the DTM ratio as a function of redshift, both considering the DTM derived from the visual extinction $A_V$ and the total metal column density, DTM$_{\rm SED}$, and the depletion-derived DTM$_{\rm mass}$. We observe no clear evolution of DTM$_{\rm SED}$ with redshift, with a Pearson correlation coefficient of $r=-0.01$, and the sample overall shows a large scatter. This is consistent with earlier results using GRB and quasar absorbers to probe DTM$_{\rm SED}$ \citep{Zafar13}. The depletion-derived DTM$_{\rm mass}$ shows a mild evolution with redshift consistent with simulations (e.g. \citealt{Li19}), with a best-fit ${\rm DTM} = (-0.03\pm 0.01)\times z + 0.35\pm 0.05$, and a Pearson $r$ coefficient of $r=-0.34$. The GRB-selected galaxies further show systematically lower DTM$_{\rm mass}$ than the Milky Way (MW), Lyman-break galaxies at $z\approx 3$ \citep{Shapley20}, and are on average also more dust deficient than the Small and Large Magellanic Clouds (SMC and LMC) (Konstantopoulou et al., subm.). Previous studies of the DTM ratios of high-redshift galaxies using GRB absorbers reached similar conclusions, though based on a different parametrization of the DTM relative to the Milky Way average \citep{DeCia13,Wiseman17}. 

\begin{figure}[!t]
    \centering
    \includegraphics[width=9.2cm]{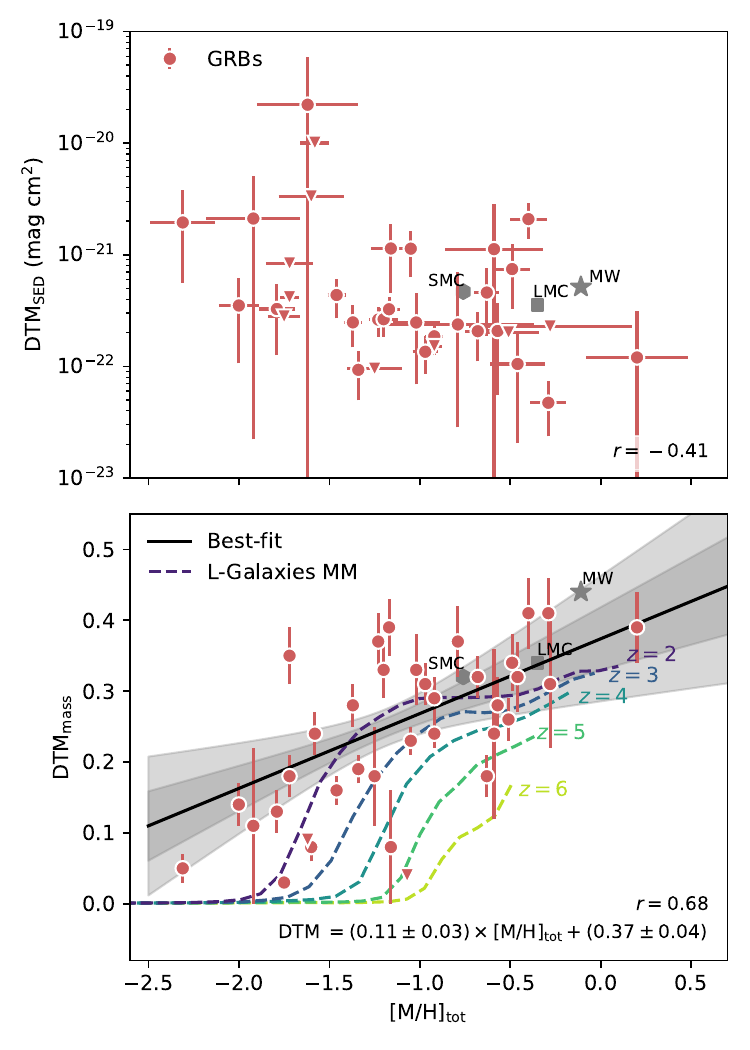}
    \caption{SED-derived dust-to-metals ratio, DTM$_{\rm SED} = \log A_V - (\log N_{\rm HI}+{\rm [M/H]_{\rm tot}})$ (top) and depletion-derived DTM$_{\rm mass}$ ratio (bottom) as a function of dust-corrected metallicity ${\rm [M/H]_{\rm tot}}$. The symbol notation follows Fig.~\ref{fig:DTMz}. Median relations at five discrete redshifts are also shown from the L-Galaxies MM simulation (Yates et al. in prep.). We observe a potential anti-correlation of DTM$_{\rm SED}$ with metallicity, but a stronger, positive correlation of DTM$_{\rm mass}$ with increasing metallicity.}
    \label{fig:DTMmet}
\end{figure}

In Fig.~\ref{fig:DTMmet} we now consider the evolution of the DTM ratio as a function of the total dust-corrected metallicity, again using both the extinction and depletion-derived expressions for the DTM. We observe a large scatter in the relation with DTM$_{\rm SED}$, but find evidence for a potential mild anti-correlation with increasing metallicity with $r = -0.41$. On the contrary, we observe a significant correlation with $r = 0.68$ of DTM$_{\rm mass}$ with increasing metallicity, with a best-fit relation of ${\rm DTM_{mass}} = (0.11\pm 0.03)\times {\rm [M/H]_{tot}} + (0.37\pm 0.04)$. This suggests that galaxies with metallicities relative to solar of $10\%$ to $1\%$ will have DTM$_{\rm mass}$ ratios that are $\approx 60\%$ to $\approx 30\%$ of the Galactic average. This is in good qualitative agreement with predictions from some simulations \citep{Vijayan19,Hou19,Graziani20}. However, GRB hosts do suggest more efficient dust production at low metallicities above $z\sim{}2$ than is found in simulations such as L-Galaxies MM (see Fig.~\ref{fig:DTMmet} and Yates et al. in prep.). This is likely due to inefficient grain growth (or other production mechanisms) at high redshift in such simulations, although biases in observational samples could also play a role (see further discussion in Sect.~\ref{ssec:dustbias}). We also caution that absorbing gas is a mix of clouds with different chemical properties, and this complexity might be difficult to take into account in the simulations.

The discrepancy between the metallicity evolution of DTM$_{\rm SED}$ and DTM$_{\rm mass}$ might originate from the distinct dust-phases probed by these two approaches. 
Dust depletion (and thus DTM$_{\rm mass}$) traces the amount of dust in the warm neutral medium in the GRB host galaxy ISM/CGM and is less sensitive to clumps of dense cold gas, in particular if rich in carbonaceous grains (Konstantopoulou et al. subm.). $A_V$ on the other hand probes the integrated extinction along the line of sight and is therefore less sensitive to more more diffuse and dust-poor regions in the GRB host galaxy, as well as the presence of large grains that may produce grey extinction.  
Based on our results, there is evidence for DTM$_{\rm mass}$ to be more tightly linked to the total metallicity of the star-forming host galaxy. The evolution with redshift is thus likely just a consequence of the DTM-[M/H]$_{\rm tot}$ relation and the overall chemical enrichment of star-forming galaxies with redshift. 
We caution that our result on the tentative anti-correlation of DTM$_{\rm SED}$ with metallicity is likely nonphysical as dust growth in the ISM is generally expected to result in an increasing DTM with metallicity. If no ISM dust growth is considered, such that the dust would purely originate from stellar sources, the DTM should be constant but not decrease with metallicity \citep{Mattsson14}. 


\begin{table}\label{tab:dtmdtg}
{\renewcommand{\arraystretch}{1.3}}
\centering
\caption{Depletion-derived dust-to-metal (DTM) and dust-to-gas (DTG) mass ratios.}
	 \begin{tabular}{lccc}
	 	 \hline
	 	 \multicolumn{1}{c}{GRB}       & $\text{z}_{\text{GRB}}$   & DTM$_{\rm mass}$   & DTG$_{\rm mass}$   \\
	 	 \hline
 000926 &   $2.0380$  & $0.39^{+0.05}_{-0.05}$ &  $(8.38^{+5.50}_{-5.50}) \times 10^{-3} $ \\
 030226 &   $1.9870$  & $<0.04$ &  $<4.6\times 10^{-5} $ \\
 050730 &   $3.9690$  & $0.05^{+0.02}_{-0.02}$ &  $(3.50^{+1.80}_{-1.80}) \times 10^{-6} $ \\
 050820A &   $2.6150$  & $0.34^{+0.04}_{-0.04}$ &  $(1.49^{+0.38}_{-0.38}) \times 10^{-3} $ \\
 050922C &   $2.1990$  & $0.11^{+0.11}_{-0.11}$ &  $(1.78^{+2.13}_{-2.07}) \times 10^{-5} $ \\
 071031 &   $2.6920$  & $0.03^{+0.01}_{-0.01}$ &  $(6.70^{+3.10}_{-3.10}) \times 10^{-6} $ \\
 080413A &   $2.4330$  & $0.08^{+0.02}_{-0.02}$ &  $(2.79^{+1.35}_{-1.35}) \times 10^{-5} $ \\
 081008 &   $1.9685$  & $0.26^{+0.03}_{-0.03}$ &  $(1.09^{+0.44}_{-0.44}) \times 10^{-3} $ \\
 090809A &   $2.7373$  & $0.32^{+0.05}_{-0.05}$ &  $(1.50^{+0.56}_{-0.56}) \times 10^{-3} $ \\
 090926A &   $2.1069$  & $0.35^{+0.04}_{-0.04}$ &  $(8.82^{+1.39}_{-1.39}) \times 10^{-5} $ \\
 100219A &   $4.6676$  & $0.08^{+0.08}_{-0.08}$ &  $(7.15^{+7.81}_{-7.37}) \times 10^{-5} $ \\
 111008A &   $4.9910$  & $0.13^{+0.03}_{-0.03}$ &  $(2.85^{+0.87}_{-0.87}) \times 10^{-5} $ \\
 111107A &   $2.8930$  & $0.31^{+0.09}_{-0.09}$ &  $(2.17^{+2.34}_{-2.34}) \times 10^{-3} $ \\
 120119A &   $1.7285$  & $0.37^{+0.05}_{-0.05}$ &  $(7.97^{+7.79}_{-7.79}) \times 10^{-4} $ \\
 120327A &   $2.8143$  & $0.19^{+0.02}_{-0.02}$ &  $(1.14^{+0.13}_{-0.13}) \times 10^{-4} $ \\
 120716A &   $2.4874$  & $0.28^{+0.04}_{-0.04}$ &  $(1.01^{+0.23}_{-0.23}) \times 10^{-3} $ \\
 120815A &   $2.3582$  & $0.37^{+0.04}_{-0.04}$ &  $(2.93^{+0.38}_{-0.38}) \times 10^{-4} $ \\
 120909A &   $3.9290$  & $0.41^{+0.05}_{-0.05}$ &  $(2.83^{+0.73}_{-0.73}) \times 10^{-3} $ \\
 121024A &   $2.3005$  & $0.32^{+0.03}_{-0.03}$ &  $(8.88^{+1.73}_{-1.73}) \times 10^{-4} $ \\
 130408A &   $3.7579$  & $0.16^{+0.02}_{-0.02}$ &  $(7.63^{+1.33}_{-1.33}) \times 10^{-5} $ \\
 130606A &   $5.9127$  & $0.24^{+0.03}_{-0.03}$ &  $(8.59^{+1.90}_{-1.90}) \times 10^{-5} $ \\
 140311A &   $4.9550$  & $0.14^{+0.03}_{-0.03}$ &  $(1.82^{+0.60}_{-0.60}) \times 10^{-5} $ \\
 141028A &   $2.3333$  & $<0.09$ &  $<2.8 \times 10^{-5} $ \\
 141109A &   $2.9940$  & $0.28^{+0.03}_{-0.03}$ &  $(1.62^{+0.25}_{-0.25}) \times 10^{-4} $ \\
 150403A &   $2.0571$  & $0.24^{+0.02}_{-0.02}$ &  $(3.81^{+0.59}_{-0.59}) \times 10^{-4} $ \\
 151021A &   $2.3297$  & $0.31^{+0.03}_{-0.03}$ &  $(4.40^{+0.86}_{-0.86}) \times 10^{-4} $ \\
 151027B &   $4.0650$  & $0.24^{+0.12}_{-0.12}$ &  $(8.39^{+6.56}_{-6.56}) \times 10^{-4} $ \\
 160203A &   $3.5187$  & $0.29^{+0.03}_{-0.03}$ &  $(4.61^{+0.64}_{-0.64}) \times 10^{-4} $ \\
 161023A &   $2.7100$  & $0.23^{+0.02}_{-0.02}$ &  $(2.69^{+0.37}_{-0.37}) \times 10^{-4} $ \\
 170202A &   $3.6456$  & $0.33^{+0.05}_{-0.05}$ &  $(4.16^{+1.39}_{-1.39}) \times 10^{-4} $ \\
 181020A &   $2.9379$  & $0.33^{+0.04}_{-0.04}$ &  $(2.75^{+0.61}_{-0.61}) \times 10^{-4} $ \\
 190114A &   $3.3764$  & $0.39^{+0.04}_{-0.04}$ &  $(3.58^{+0.64}_{-0.64}) \times 10^{-4} $ \\
 190106A &   $1.8599$  & $0.41^{+0.05}_{-0.05}$ &  $(2.18^{+0.56}_{-0.56}) \times 10^{-3} $ \\
 190919B &   $3.2241$  & $0.18^{+0.07}_{-0.07}$ &  $(1.37^{+0.71}_{-0.71}) \times 10^{-4} $ \\
 191011A &   $1.7204$  & $0.18^{+0.03}_{-0.03}$ &  $(5.71^{+1.22}_{-1.22}) \times 10^{-4} $ \\
 210905A &   $6.3118$  & $0.18^{+0.03}_{-0.03}$ &  $(4.64^{+1.53}_{-1.53}) \times 10^{-5} $ \\
	 	 \hline
	 \end{tabular}
\end{table}

\subsection{The evolution of the dust-to-gas ratio}

\begin{figure}
    \centering
    \includegraphics[width=9cm]{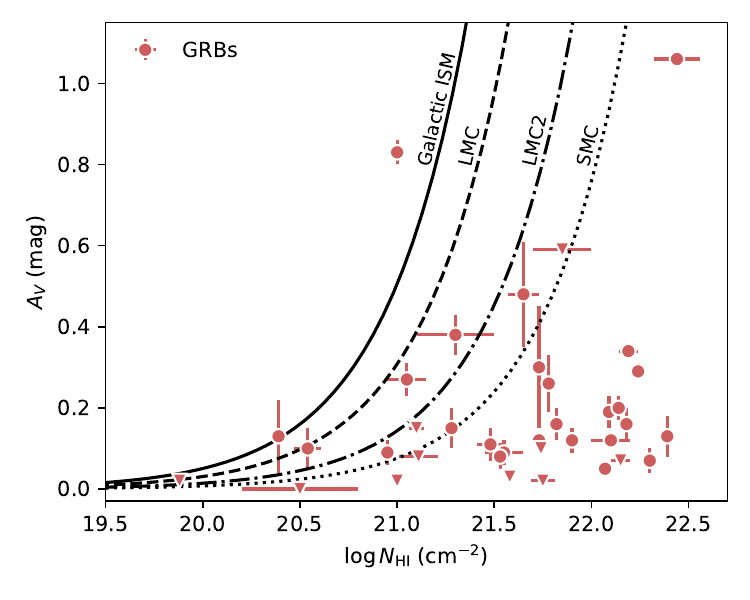}
    \caption{Visual extinction $A_V$ vs. \hi\ column density $N_{\rm HI}$, i.e. the dust-to-gas (DTG) ratio. The symbol notation follows Fig.~\ref{fig:DTMz}. For comparison are overplotted the average DTG ratios from specific sightlines in the Local Group (MW, LMC, LMC2 and SMC) from \citet{Gordon03}. GRBs typically probe sightlines with lower $A_V$ for a given $N_{\rm HI}$ than the local galaxies.}
    \label{fig:avnhi}
\end{figure}

The \hi\ column densities and visual extinction, $A_V$, measured directly in the GRB sightlines provide an independent measure of the DTG ratio, DTG$_{\rm SED}$ = $A_V/N_{\rm HI}$, in high-redshift galaxies. In Fig.~\ref{fig:avnhi} we show the distribution of $A_V$ and $N_{\rm HI}$ observed for the GRB sample. For comparison, the average DTG$_{\rm SED}$ ratios from specific Galactic sightlines and toward the SMC bar, the mean LMC and the LMC2 supershell from \citet{Gordon03} are shown as well. We find that the majority of the GRB sightlines probe DTG ratios lower than observed in these local galaxies, and measure an average value $(A_V/N_{\rm HI})_{\rm GRB}$ = $8.95\times 10^{-23}$\,mag\,cm$^{2}$. This is consistent with previous estimates of high-$z$ GRB \citep{Schady10,Zafar11} and general quasar DLA \citep{Vladilo08,Khare12} sightlines. We note that the molecular hydrogen gas fraction is $\approx 5\%$ at maximum in the GRB absorption systems \citep{Bolmer19,Heintz19c}, and is therefore negligible in the derivation of the DTG.

Similar to DTM$_{\rm mass}$, we can also infer the mass of an element X in the dust-phase relative to the total gas mass, the DTG mass ratio, DTG$_{\rm mass}$, here based on Eq.~\ref{eq:dtm} as
\begin{equation}
    {\rm DTG_{mass}} = \frac{M_{\rm dust}}{M_{\rm gas}} = {\rm DTM} \times 10^{\rm [M/H]_{\rm tot}} \times Z_\odot
\end{equation}
where [M/H]$_{\rm tot}$ is again the total dust-corrected metallicity and $Z_\odot=0.0139$ is the solar metallicity by mass \citep{Asplund21}. The derived DTG$_{\rm mass}$ ratios span $10^{-5}- 2\times 10^{-3}$ as summarized in Table~2. 

\begin{figure}
    \centering
    \includegraphics[width=9cm]{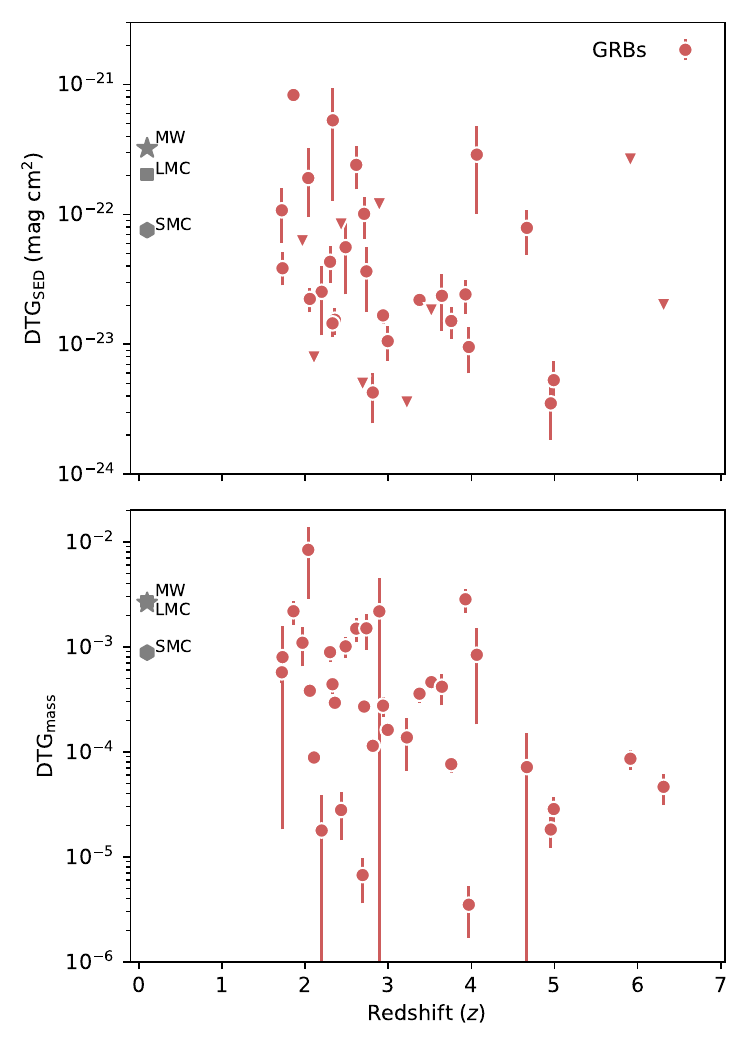}
    \caption{SED-derived dust-to-gas ratio, DTG$_{\rm SED} = A_V / N_{\rm HI}$ (top) and depletion-derived DTG$_{\rm mass}$ (bottom) as a function of redshift. The symbol notation follows Fig.~\ref{fig:DTMz}. For comparison we mark the DTG ratios of the MW, LMC, and SMC. The GRB sightlines through high-redshift galaxies typically probe lower DTG content than inferred from these local galaxies.}
    \label{fig:DTGz}
\end{figure}

To explore the cause of the low relative dust to gas content in the GRB sightlines, we first examine the evolution of the DTG as a function of redshift in Fig.~\ref{fig:DTGz}. Here, we again consider the DTG ratios calculated both from the SED fit of the dust extinction DTG$_{\rm SED}$ and the depletion-derived DTG$_{\rm mass}$. We observe a tendency for a decreasing DTG as a function of redshift in both parameterizations albeit with a large scatter. This implies that the difference in the SED- and dust depletion-based DTM is likely not due to the differences in how the dust is probed. Notably, the bulk of the GRB-selected galaxies at $z>2$ have DTG$_{\rm mass}$ measurements below the MW average, reaching three orders of magnitudes lower at DTG = ${\rm DTG_{mass}} = 3.5\times 10^{-6}$ (GRB\,050730) (see also \citealt{Wiseman17}). 

\begin{figure}
    \centering
    \includegraphics[width=9cm]{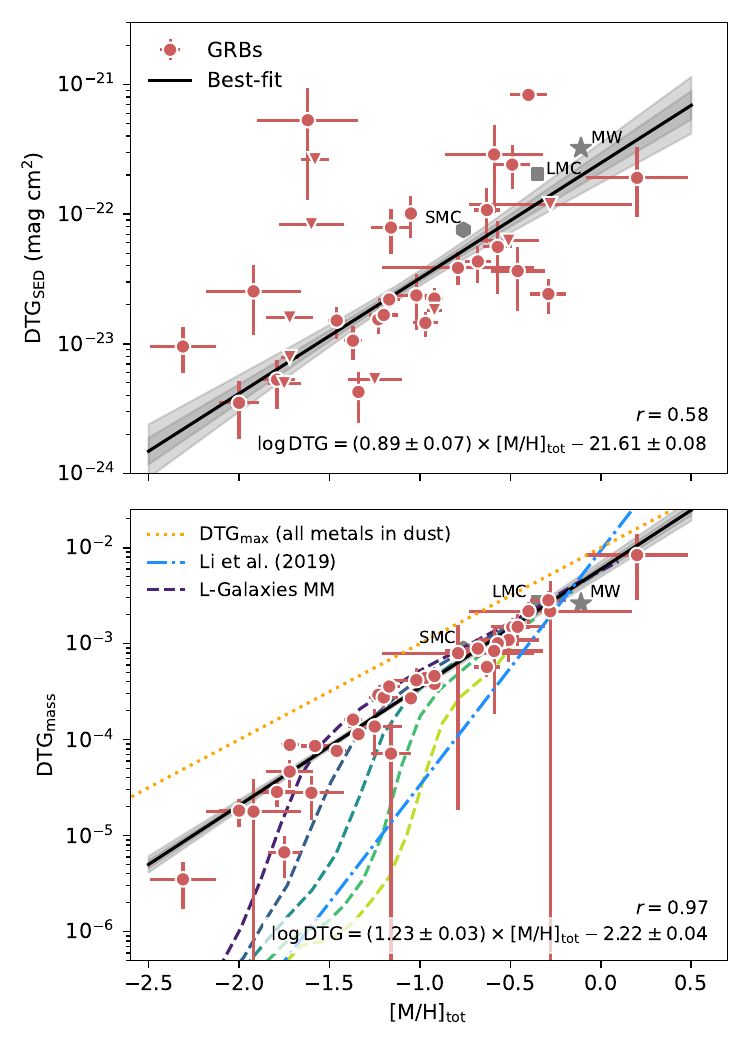}
    \caption{SED-derived dust-to-gas ratio, DTG$_{\rm SED} = A_V / N_{\rm HI}$ (top) and depletion-derived DTG$_{\rm mass}$ (bottom) as a function of dust-corrected metallicity ${\rm [M/H]_{\rm tot}}$. The symbol notation follows Fig.~\ref{fig:DTMz}. We find strong correlations between the DTG inferred both from the SED (top panel) and from depletion (bottom panel) with the metallicity. In the bottom panel we also show the predictions from the {\tt Simba} simulation at $z\sim 0-6$ \citep[blue dotted line;][]{Li19} and the L-Galaxies MM simulation at $z\sim{}2-6$ (dashed lines; Yates et al. in prep.). The orange dashed line is the limit in which all metals are incorporated into grains (${\rm DTM} = 1$).} 
    \label{fig:DTGmet}
\end{figure}

In Fig.~\ref{fig:DTGmet} we further examine the evolution of the DTG as a function of metallicity. Here we also plot the DTG ratios measured for the MW, SMC, and LMC. We find strong correlations between the DTG ratios inferred both from the dust extinction and from the depletion-derived mass ratio, with best-fit relations
\begin{equation}
    \log {\rm DTG}_{\rm SED} = (0.89\pm 0.07)\times {\rm [M/H]_{tot}} - (21.61\pm 0.08)\, {\rm mag\,cm^{-2}}
\end{equation}
assuming the DTG ratio described by $A_V/N_{\rm HI}$, with a Pearson correlation coefficient of $r=0.58$, and
\begin{equation}
    \log {\rm DTG_{mass}} = (0.92\pm 0.06)\times {\rm [M/H]_{tot}} - (2.81\pm 0.08)
\end{equation}
assuming DTG$_{\rm mass}$ measured from the depletion level, with a correlation coefficient of $r=0.97$. These relations are in good agreement with the measurements of the DTG ratios in the MW, SMC, and LMC, populating the high-metallicity end. Notably, GRB\,141028A have a DTG$_{\rm SED}$ ratio of $A_V/N_{\rm HI} \gtrsim 5\times 10^{-22}$\,mag\,cm$^2$, exceeding the average MW DTG ratio of $4.5\times 10^{-22}$\,mag\,cm$^2$ \citep{Watson11}, but at substantially lower metallicities of $-1.62\pm 0.28$. On the other hand, the depletion-derived DTG$_{\rm mass}$ is observed to follow a tight correlation with the dust-corrected metallicity [M/H]$_{\rm tot}$. This suggests that for some particular sightlines like in GRBs\,141028A, the extinction in the line-of-sight caused by dust grains are substantially larger than predicted from the overall depletion strength \citep[see also][]{Savaglio04,Wiseman17,Bolmer19,Konstantopoulou22}. This could potentially indicate that a dominant contribution from dust grains probed via the extinction are not recovered in the depletion analysis. Potential causes of the discrepancy are that a substantial amount of dust in this system is either confined in clumps of cold neutral medium, or in intervening systems along the line of sight.

We further compare our relations to the predictions by \citet{Li19} based on the {\tt Simba} cosmological hydrodynamic galaxy formation simulation and the L-Galaxies simulations by Yates et al. (in prep.). Although L-Galaxies MM matches the GRB relation well for systems below $z\sim{}3$, both simulations find steeper slopes for the metallicity evolution of the DTG than observed in the higher-redshift GRB sightlines. This could potentially indicate a more efficient dust production in the low-metallicity regime probed by these high-redshift sightlines than what is currently prescribed in the simulations, as also indicated previously by the excess DTM$_{\rm mass}$ ratios. 

\subsection{Quantifying the dust bias in our sample} \label{ssec:dustbias}

\begin{figure}
    \centering
    \includegraphics[width=9cm]{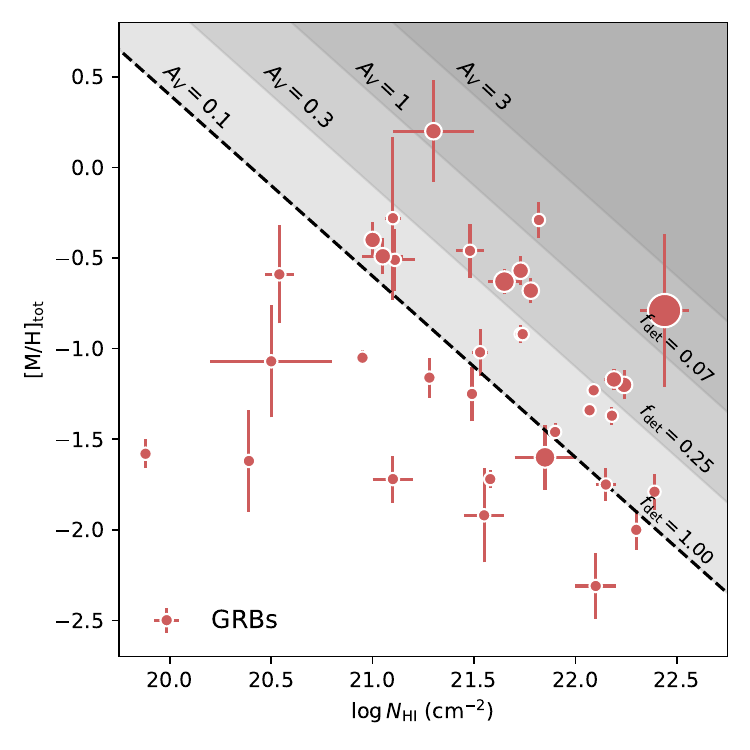}
    \caption{Dust-corrected metallicity as a function of \hi\ column density. The red dots again present the main GRB sample, but here size-coded as a function of the SED-derived $A_V$. The grey-shaded regions represent increasing $A_V$, marked for each region, assuming the constant DTM$_{\rm SED}$ ratio derived in this work of $\log A_V - [\log N_{\rm HI} + {\rm [M/H]_{tot}}] = -21.4$. The estimated detection probability for GRBs in each of these $A_V$ ranges are marked as well (see Sect.~\ref{ssec:dustbias} for further details).}
    \label{fig:dustbias}
\end{figure}

Since this analysis is based on GRB afterglows observed with medium to high resolution spectroscopy, we might be biased against the most metal- or dust-rich sightlines that obscures the afterglow light below the detection threshold \citep{Ledoux09}. To investigate whether our parent sample is subject to this bias, we compare the $A_V$ distribution to that of the more unbiased photometric GRB sample presented by \citet{Covino13}, limited to $z>1.7$. Following \citet{Heintz19c}, we normalize the two distributions by the number of bursts at $A_V<0.1$\,mag (assuming that the spectroscopic sample is at least complete to this limit) and then compute the detection probability, $f_{\rm det}$, of the fraction of GRBs in the spectroscopic sample versus that of the unbiased sample at the given range in $A_V$ (see Fig.~\ref{fig:dustbias}). We find that the spectroscopic sample is complete up to $A_V = 0.3$\,mag (i.e. $f_{\rm det}=1$), but that we are only recovering 25\% of the expected GRB sightlines at $A_V = 0.3-1.0$\,mag ($f_{\rm det}=0.25$). At $A_V > 1$\,mag, the spectroscopic sample is only 7\% complete and even less at larger visual extinctions. 

An additional complication might be introduced via our selection. For instance, since we require a detection of at least a set of metal lines at $3\sigma$ this will inherently disfavor low-metallicity systems but less severely potentially dust extinguished sightlines due to the broader spectral continuum range observed. This might partly alleviate the tension of our observations with the simulations in the low-metallicity regimes of the DTG and DTM mass ratios (Figs.~\ref{fig:DTMmet} and \ref{fig:DTGmet}), and potentially indicate even steeper correlations with metallicity. 

Conservatively, we can thus only conclude that the relations derived here are representative of the moderately extinguished ($A_V \lesssim 0.3$\,mag) GRB host-galaxy population, and may deviate at larger dust columns. We note, however, that the large majority of the GRB population in complete samples of GRBs has $A_V<1$\,mag, and there is no clear evidence that we are missing the most dust-obscured bursts at $z>2$ \citep{Kruhler11,Kruhler12}. Nevertheless, while this potential dust obscuration bias may limit the number of the most dust- and metal-rich GRBs in the spectroscopic sample, the inferred trends of the relative DTM and DTG ratios are likely still valid. 
The main advantage of using GRBs as cosmic probes, lie in the high-$z$, low-metallicity regime, which is more difficult and time consuming to probe with direct emission-based surveys. 

\section{Summary and Future Outlook} \label{sec:conc}

In this work, we have presented the most comprehensive analysis to date of the chemical enrichment and the evolution of dust and metals in the ISM of star-forming galaxies at $z= 1.7-6.3$ hosting GRBs. We have compiled all GRB afterglow spectra observed through more than two decades (years $2000-2021$) with sufficient spectral resolution ($\mathcal{R}>7000$) and S/N $>3$ per resolution bin to enable robust measurements of the element abundances in the GRB lines of sight. GRBs are particularly efficient and relatively unbiased tracers of dense star-forming regions and thereby provide a unique view into the star-forming ISM properties of high-redshift galaxies in absorption that are otherwise difficult to probe in emission. 

We found that the GRB-selected, star-forming galaxies had metallicities, corrected for the abundance of elements in the dust-phase, that were on average evolving as a function of redshift following ${\rm [M/H]}_{\rm tot}(z) = (-0.21\pm 0.04)z - (0.47\pm 0.14)$. These galaxies revealed a slower gradual metal build-up compared to DLAs in quasar sightlines \citep{DeCia18}, which traces the neutral gas on larger scales around galaxies. Our observations further exhibited a large scatter in the dust-corrected metallicities at a given redshift, which is not captured in most state-of-the-art galaxy evolution simulations, although there is overall agreement with the chemical enrichment as a function of cosmic time. The largest observational uncertainty in this relation is reflected by the more sparse population of GRB afterglows detected at the highest redshifts, $z\gtrsim 5$. This particular population may also be subject to a more severe selection bias than at lower redshifts since they will appear optically ``dark''.

Based on our observations, we further derived the redshift and metallicity evolution of the DTG and DTM ratios in GRB-selected, star-forming galaxies at $z>2$. Previously, the far-infrared emission of galaxies have been used to determine the mass and temperature of the dust in the ISM \citep[e.g.,][]{Draine07}. These dust mass estimates have commonly been used to infer the total gas or ISM mass of high-redshift galaxies \citep[e.g.,][]{Magdis12,Scoville16}, but typically assuming average MW dust-to-gas ratios or conversion factors. However, these might be systematically underestimated if the fraction of dust relative to the total gas and metal abundance in the high-redshift, metal-poor regimes of galaxies changes substantially. Indeed, in this work we observed that all the $z\gtrsim 2$ GRB-selected galaxies probe sightlines with lower DTM and DTG values compared to the Galactic average. In particular, we found that the average DTM mass ratios at $z\approx 2$ and $z\approx 6$ were 0.3 and 0.15, respectively, $2-3\times$ lower than observed in the Milky Way. Similarly, the average gas-to-dust mass ratio is $\approx 150$ for the MW, where, for comparison, we observed sight lines reaching gas-to-dust mass ratios of $\approx 10^{5}$ at $z\gtrsim 3$ or metallicities [M/H]$_{\rm tot} < -1.5$ (i.e. 3\% solar). 

The DTM and DTG mass ratios derived here do not rely on any conversion factors, and thus provide an accurate measure of the relative mass fractions of gas, dust, and metals along the GRB line of sight. However, since GRBs only probe narrow pencil-beam sightlines through their host galaxies, the total integrated dust and metal abundances are difficult to determine without knowing the size and morphology of the systems. For example, the distribution of dust, metals and gas within the galaxy ISM may impact the measured DTM and DTG along a single line-of-sight. In spatially resolved observations, the DTG mass ratio has been found to vary as a function of hydrogen surface density \citep{Clark23} and metallicity \citep{Solis20}, with variations of over an order of magnitude within a single galaxy. A similar variation is also seen as a function of radius in galaxy evolution simulations out to $4-5~R_{\rm e}$ (Yates et al. in prep.). The DTM mass ratio is similarly observed to vary within galaxies as a function of metallicity \citep{chiang18}, irrespective of the $\alpha_{\rm CO}$ conversion factor applied. The average DTM and DTG mass ratios measured along single sightlines in absorption may thus be weighted differently to the average properties probed in emission. For example, the latter method would likely be more biased towards the higher metallicity and higher surface density regions of a galaxy, that have higher DTM and DTG mass ratios. It is also important to consider differences in the gas and dust radial profiles, with the former generally extending out to larger radii by up to 50-100\% \citep{Thomas04}. This could dilute the DTG and DTM measured along GRB sightlines, although this is not supported by simulations which predict systematically lower DTG and DTM mass ratios than measured along GRB sight lines (e.g. Figs.~\ref{fig:DTMmet} and \ref{fig:DTGmet}). Some additional caveats in directly mapping the GRB results to simulations and emission-selected galaxy studies also include the physical environments of the GRBs, such as the ionization state and density of the gas.


Despite the expected differences between absorption and emission probes, the GRB absorption approach has proven to be extremely effective in determining the [\ci]-to-H$_2$ conversion factor in high-redshift, low-metallicity galaxies \citep{Heintz20}, which is otherwise difficult to constrain \citep{Bolatto13}, and even provide novel constraints on the \hi\ gas masses of galaxies at $z>2$ \citep{Heintz21,Heintz22}. This would suggest that GRBs are able to probe galaxy average properties in the radial direction, even if only along a single sightline. The DTM and DTG mass abundance ratios derived here, therefore like enable more accurate determinations of the total gas or ISM masses of high-redshift galaxies, based on the dust masses inferred from the far-infrared dust continuum emission of independent galaxy samples. 

In the near-future, this field is certain to rapidly advance with the new spectroscopic observations of high-redshift galaxies with the JWST. Already now, gas-phase metallicities of galaxies have been measured using temperature-sensitive diagnostics up to $z\approx 9$ in emission \citep{Heintz22met,Curti23,Nakajima23,Sanders23}, and even inferred approximately through strong-line diagnostics at $z>10$ \citep{Bunker23,Hsiao23,Heintz23lya}. The unique synergy between JWST and the Atacama Large Millimetre/submillimetre Array (ALMA) further enables direct DTM and DTG mass ratio measurements of galaxies well into the epoch of reionization at $z>6$ \citep{Heintz23}. Characterizing the host galaxies of GRBs in emission with similar observations would be the natural next step, solidifying the link between galaxy properties derived in absorption and emission.

\begin{acknowledgements}
First and foremost, we would like to thank all the astronomers around the globe that has been on duty throughout the years, tirelessly following up GRBs, which resulted in this exquisite GRB spectroscopic afterglow legacy sample. 
K.E.H. acknowledges support from the Carlsberg Foundation Reintegration Fellowship Grant CF21-0103.
The Cosmic Dawn Center (DAWN) is funded by the Danish National Research Foundation under grant No. 140. 
A.D.C. and C.K. acknowledge support by the Swiss National Science Foundation under grant 185692. AR acknowledges support from the INAF project Premiale Supporto Arizona \& Italia.
A.S. and S.D.V. acknowledge support from CNES and DIM-ACAV+.
G.S. acknowledges the support by the State of Hesse within the Research Cluster ELEMENTS (Project ID 500/10.006)
Based on observations collected at the European Organisation for Astronomical Research in the Southern Hemisphere.
\end{acknowledgements}

\bibliographystyle{aa}
\bibliography{ref}

\clearpage


\appendix
\section{Column densities and visual extinctions of new GRB afterglows} \label{sec:app}

Here we present the measurements and detail the derivations of the metal abundances and line-of-sight visual extinction for the three new GRBs examined in this work: GRBs\,190106A, 190919B, and 191011A. 

\subsection{\texorpdfstring{GRB\,190106A}{GRB 190106A}} \label{ssec:a1}

\begin{figure}[!h]
    \centering
    \includegraphics[width=9cm]{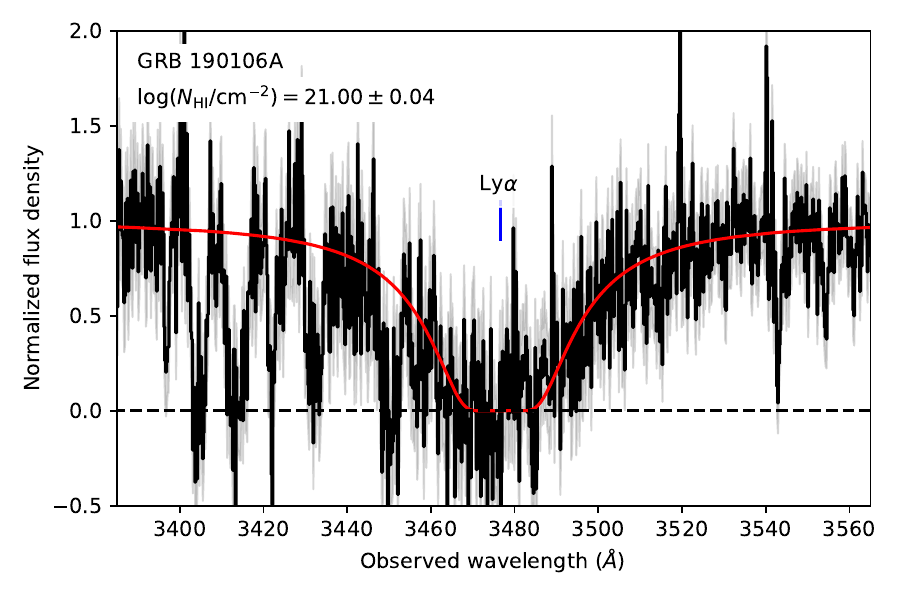}
    \caption{VLT/X-shooter GRB afterglow spectrum of GRB\,190106A. The normalized 1D spectrum is shown in black, with the associated error spectrum shown by the grey region. The best-fit damped Lyman-$\alpha$ profile with $\log(N_{\rm HI}/{\rm cm}^{-2}) = 21.00\pm 0.04$ is shown in red.}
    \label{afig:grb190106a_hi}
\end{figure}

\begin{figure}[!h]
    \centering
    \includegraphics[width=9cm]{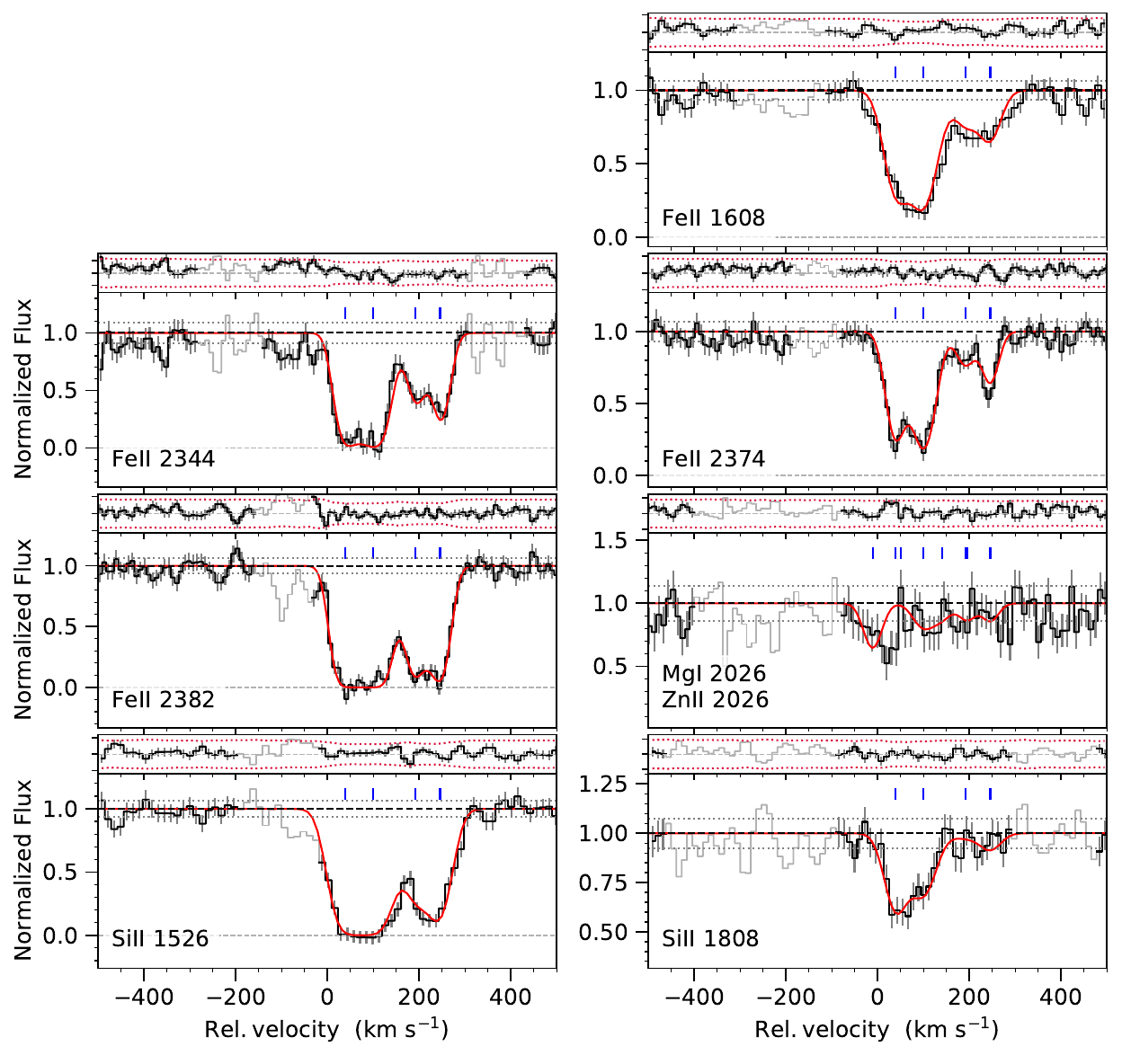}
    \caption{Metal absorption-line modelling of GRB\,190106A. The normalized VLT/X-shooter spectrum is shown in black, with the best-fit Voigt profiles to each of the line transitions and velocity components (marked in blue) shown by the red curves. At the top of each panel are shown the residual plots.}
    \label{afig:grb190106a_met}
\end{figure}

To determine the metallicity of GRB\,190106A, we first measure the \hi\ column density based on the broad Lyman-$\alpha$ transition using {\tt VoigtFit}. The VLT/X-shooter afterglow spectrum and the best-fit model with $\log(N_{\rm HI}/{\rm cm}^{-2}) = 21.00\pm 0.04$ is shown in Fig.~\ref{afig:grb190106a_hi}. Then, we derive the metal abundances of each element X by modelling the absorption-line profiles of the low-ionization metal transitions as outlined in Sect.~\ref{sec:analysis}. GRB\,190106A shows multiple velocity components as displayed in Fig.~\ref{afig:grb190106a_met}, but we report only the sum for the Zn and Fe transitions in Table~\ref{tab:results} for the metallicity and depletion measurements. The spectroscopic redshift of the strongest low-ionization component which we consider as the redshift of the GRB is $z_{\rm GRB} = 1.8599$.

\begin{figure}[!h]
    \centering
    \includegraphics[width=9cm]{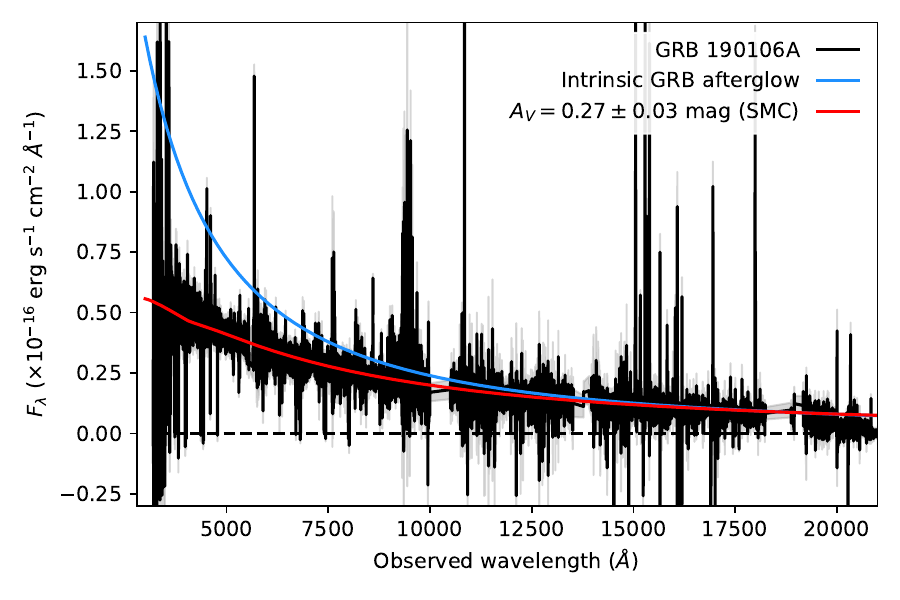}
    \caption{Visual extinction $A_V$ of GRB\,190106A. The extracted 1D VLT/X-shooter GRB afterglow spectrum is shown in black, the intrinsic power-law shape derived from the {\it Swift}/XRT photon index in blue, and the best-fit model with $A_V = 0.27\pm 0.03$\,mag (assuming an SMC extinction curve) in red.}
    \label{afig:grb190106a_av}
\end{figure}

Following the description in Sect.~\ref{ssec:avext}, we determine the visual extinction along the line of sight to GRB\,190106A using the X-ray derived photon index, $\Gamma = 1.90$, as prior on the intrinsic GRB afterglow power-law slope, $F_\lambda = F_{0}\lambda^{\Gamma - \Delta \beta - 3}$. We find that the spectral synchrotron cooling break is consistent with $\Delta = 0.5$, such that the intrinsic spectral $F_\lambda$ power-law becomes $\beta = -1.60$. Modelling the observed afterglow spectrum with an SMC extinction curve parametrization from \citet{Gordon03}, yields $A_V = 0.27\pm 0.03$\,mag (see Fig.~\ref{afig:grb190106a_av}).

\subsection{\texorpdfstring{GRB\,190919B}{GRB 190919B}}

\begin{figure}[!h]
    \centering
    \includegraphics[width=9cm]{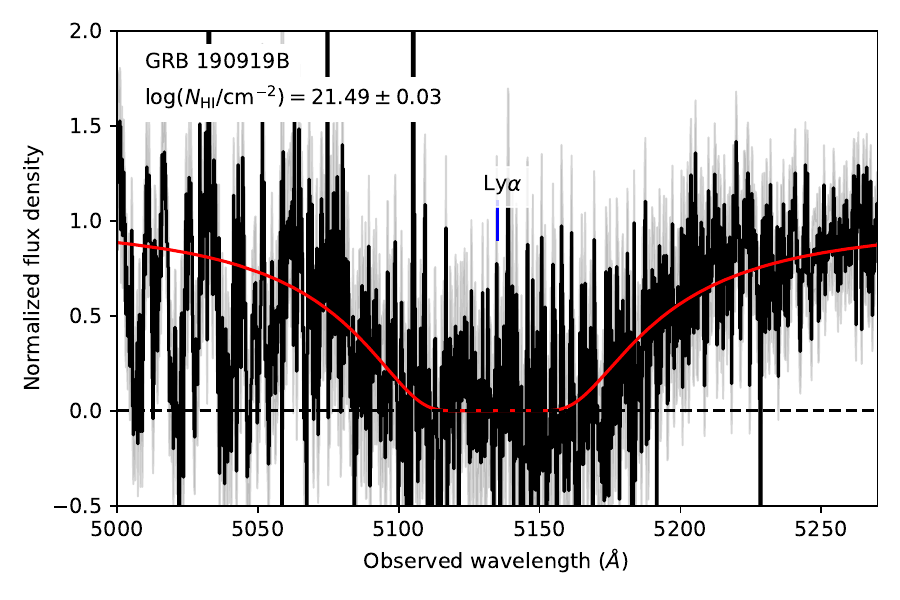}
    \caption{Same as in Fig.~\ref{afig:grb190106a_hi} but for GRB\,190919B, with a best-fit $\log(N_{\rm HI}/{\rm cm}^{-2}) = 21.49\pm 0.03$.}
    \label{afig:grb190919b_hi}
\end{figure}

\begin{figure}[!h]
    \centering
    \includegraphics[width=9cm]{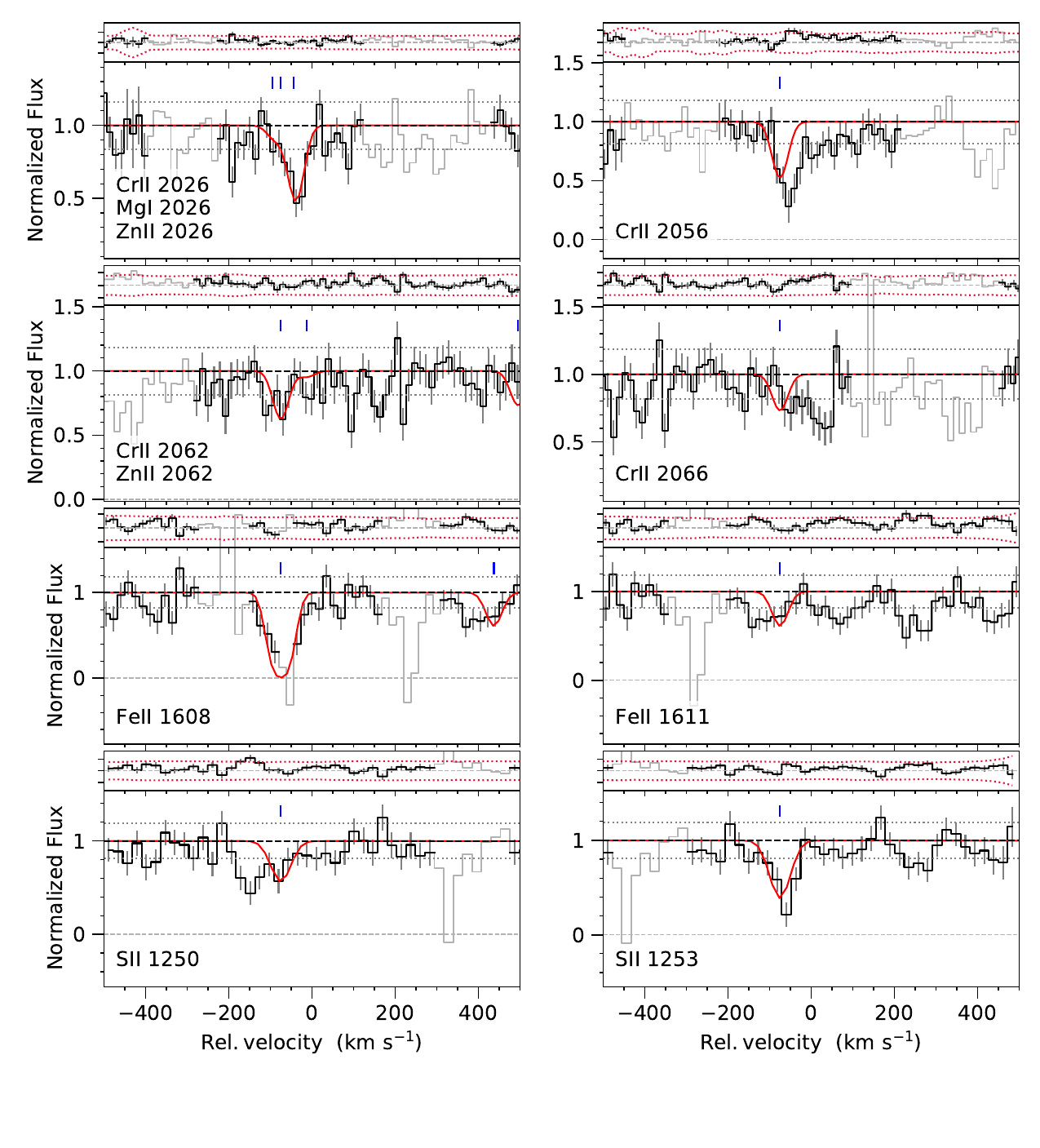}
    \caption{Same as in Fig.~\ref{afig:grb190106a_met} but for GRB\,190919B.}
    \label{afig:grb190919b_met}
\end{figure}

\begin{figure}[!h]
    \centering
    \includegraphics[width=9cm]{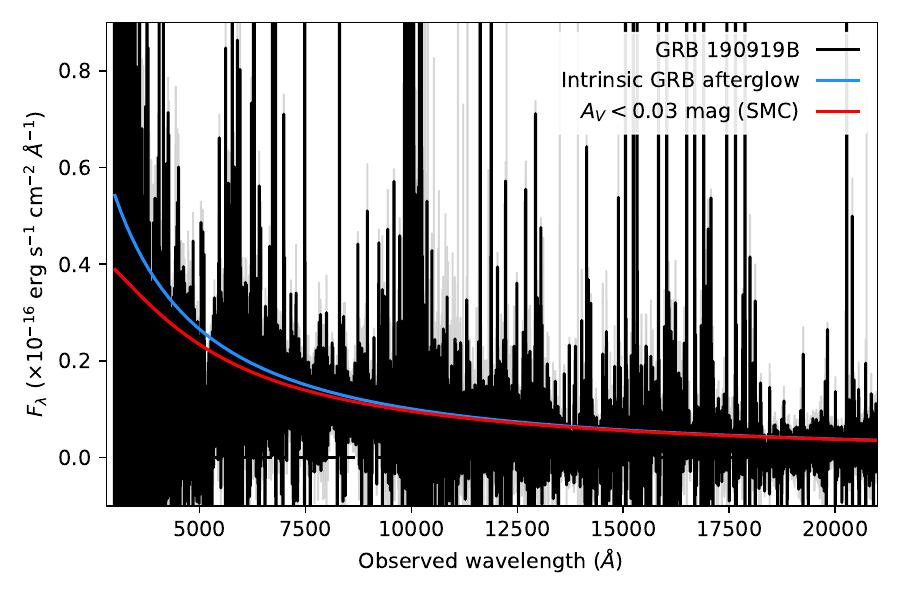}
    \caption{Same as in Fig.~\ref{afig:grb190106a_av} but for GRB\,190919B, with a constraint on $A_V < 0.03$\,mag ($1\sigma$).}
    \label{afig:grb190919b_av}
\end{figure}

Following the same procedure as in Sect.~\ref{ssec:a1} but for GRB\,190919B, we derive a best-fit \hi\ column density of $\log(N_{\rm HI}/{\rm cm}^{-2}) = 21.49\pm 0.03$ as shown in Fig.~\ref{afig:grb190919b_hi}. We determine the metal abundances from the single absorption component detected in the afterglow spectrum at $z_{\rm GRB} = 3.2241$, as reported in Table~\ref{tab:results} and shown in Fig.~\ref{afig:grb190919b_met}. The XRT photon index $\Gamma = 2.10$ represents an intrinsic optical/near-infrared power-law slope of $\beta = -1.40$, which yields a visual extinction of $A_V < 0.03$\,mag (at $1\sigma$), see Fig.~\ref{afig:grb190919b_av}. 

\subsection{\texorpdfstring{GRB\,191011A}{GRB 191011A}}

\begin{figure}[!h]
    \centering
    \includegraphics[width=9cm]{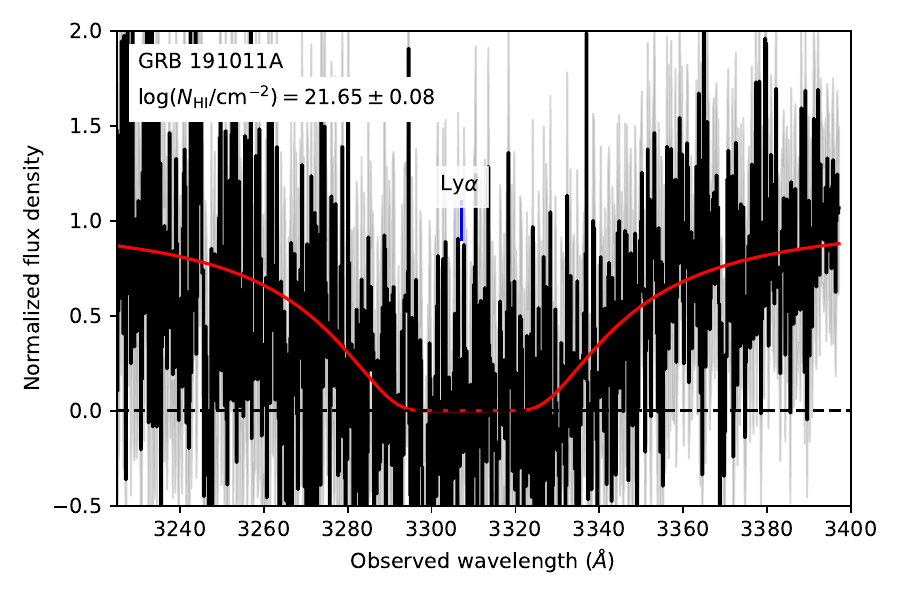}
    \caption{Same as in Fig.~\ref{afig:grb190106a_hi} but for GRB\,191011A, with a best-fit $\log(N_{\rm HI}/{\rm cm}^{-2}) = 21.49\pm 0.03$.}
    \label{afig:grb191011a_hi}
\end{figure}

\begin{figure}[!h]
    \centering
    \includegraphics[width=9cm]{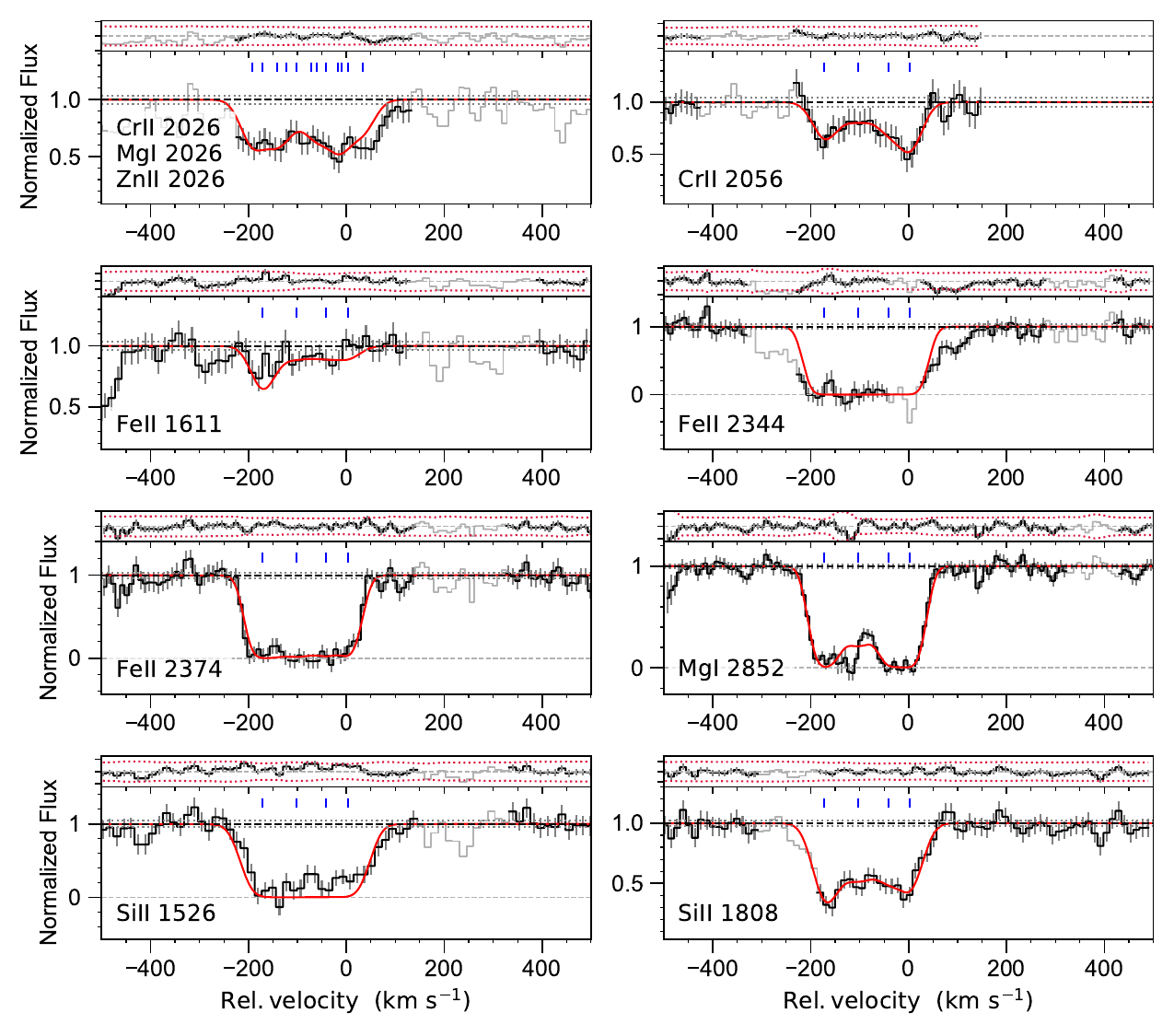}
    \caption{Same as in Fig.~\ref{afig:grb190106a_met} but for GRB\,191011A.}
    \label{afig:grb191011a_met}
\end{figure}

\begin{figure}[!h]
    \centering
    \includegraphics[width=9cm]{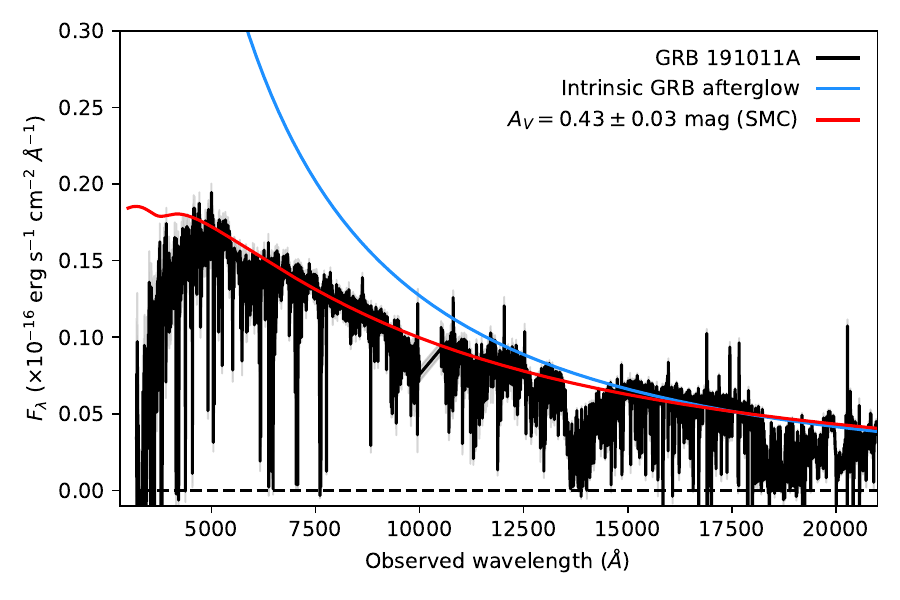}
    \caption{Same as in Fig.~\ref{afig:grb190106a_av} but for GRB\,191011A, with a best-fit of $A_V = 0.43\pm 0.03$\,mag.}
    \label{afig:grb191011a_av}
\end{figure}

For GRB\,191011A, we compute an \hi\ column density of $\log(N_{\rm HI}/{\rm cm}^{-2}) = 21.65\pm 0.08$ as shown in Fig.~\ref{afig:grb191011a_hi}. We determine the metal abundances from the sum of the velocity components detected in the afterglow spectrum, with the redshift of the strongest component being $z_{\rm GRB} = 1.7204$, as reported in Table~\ref{tab:results} and shown in Fig.~\ref{afig:grb191011a_met}. The XRT photon index $\Gamma = 1.89$ represents an intrinsic optical/near-infrared power-law slope of $\beta = -1.61$, which yields a visual extinction of $A_V < 0.43\pm0.03$\,mag, see Fig.~\ref{afig:grb191011a_av}.

\end{document}